\documentclass[fleqn,usenatbib]{mnras}

\usepackage{newtxtext,newtxmath}
\usepackage{xcolor}

\usepackage{graphicx}	% Including figure files
\usepackage{amsmath}	% Advanced maths commands

\usepackage{graphicx}	% Including figure files
\usepackage{amsmath}	% Advanced maths commands
\usepackage{xspace}
\usepackage{longtable}
\usepackage{subfig}
\usepackage{booktabs}
\usepackage{lscape}
\usepackage{siunitx}
\usepackage[symbol]{footmisc}
\usepackage[dvipsnames]{xcolor}
\usepackage{hyperref}
\usepackage{textgreek}
\usepackage{array}

\usepackage{tabularx}
\usepackage[T1]{fontenc}

\DeclareRobustCommand{\VAN}[3]{#2}
\let\VANthebibliography\thebibliography
\def\thebibliography{\DeclareRobustCommand{\VAN}[3]{##3}\VANthebibliography}

\usepackage{graphicx}	% Including figure files
\usepackage{amsmath}	% Advanced maths commands
\usepackage{xspace}
\usepackage[normalem]{ulem}
\newcommand{\planetradius}{\textcolor{black}{$R_{\rm b}=2.51\pm 0.11\,\rm R_\oplus$}\xspace}
\newcommand{\period}{\textcolor{black}{$P=60.778853^{+0.000055}_{-0.000077}\,\rm d$}\xspace}
\newcommand{\teq}{\textcolor{black}{$T_{\mathrm{eq}}=175^{+6}_{-5}\,\rm K$}\xspace}
\newcommand{\instellation}{\textcolor{black}{$S=0.22\pm 0.04\,\rm S_{\oplus}$}\xspace}
\newcommand{\radiusprecision}{\textcolor{black}{$4.5\,\%$}\xspace}
\newcommand{\planetmass}{\textcolor{black}{$M_{\rm b}=6.84\pm 0.52\,\rm M_\oplus$}\xspace}
\newcommand{\planetdensity}{\textcolor{black}{$\rho_{\rm b}=2.4\pm 0.4\,\rm g\,cm^{-3}$}\xspace}
\newcommand{\tsm}{\textcolor{black}{$\mathrm{TSM}=56\pm10$}\xspace}

\newcommand{\stellarradius}{$R_\star=0.304\pm 0.012\,\rm R_\odot$\xspace}
\newcommand{\stellarmass}{$M_\star=0.296\pm 0.026\,\rm M_\odot$\xspace}
\newcommand{\stellarteff}{$T_{\rm eff}=3215\pm 58\,\rm K$\xspace}
\title[TOI-6981: water ice-line planet]{\textcolor{black}{TOI-6981\,b: a transiting sub-Neptune at the water ice line}}
\author[M. G. Scott et al.]{
Madison G. Scott$^{\href{https://orcid.org/0009-0006-3846-4558}{\includegraphics[scale=0.5]{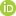}}}$,$^{1}$\thanks{E-mail: mgs947@student.bham.ac.uk (MGS)}
Amaury H.M.J. Triaud$^{\href{https://orcid.org/0000-0002-5510-8751}{\includegraphics[scale=0.5]{plots/orcid.jpg}}}$,$^{1}$
Emily K. Pass$^{\href{https://orcid.org/0000-0002-1533-9029}{\includegraphics[scale=0.5]{plots/orcid.jpg}}}$,$^{2}$ % TRES
Aleyna Adamson$^{\href{https://orcid.org/0009-0004-7125-4101}{\includegraphics[scale=0.5]{plots/orcid.jpg}}}$,$^{1}$ % kima
Karen A. Collins$^{\href{https://orcid.org/0000-0001-6588-9574}{\includegraphics[scale=0.5]{plots/orcid.jpg}}}$,$^{3}$ % LCO
\newauthor
Khalid Barkaoui$^{\href{https://orcid.org/0000-0003-1464-9276}{\includegraphics[scale=0.5]{plots/orcid.jpg}}}$,$^{4,5,6}$ % phot reduction, SPC
Allyson Bieryla$^{\href{https://orcid.org/0000-0001-6637-5401}{\includegraphics[scale=0.5]{plots/orcid.jpg}}}$,$^{3}$ % KeplerCam
Adam J.\ Burgasser$^{\href{https://orcid.org/0000-0002-6523-9536}{\includegraphics[scale=0.5]{plots/orcid.jpg}}}$,$^{7}$  % stellar char
Mark~E.~Everett$^{\href{https://orcid.org/0000-0002-0885-7215}{\includegraphics[scale=0.5]{plots/orcid.jpg}}}$,$^{8}$ % WIYN
Steve B. Howell$^{\href{https://orcid.org/0000-0002-2532-2853}{\includegraphics[scale=0.5]{plots/orcid.jpg}}}$,$^{9}$ % zorro
\newauthor
David W. Latham$^{\href{https://orcid.org/0000-0001-9911-7388}{\includegraphics[scale=0.5]{plots/orcid.jpg}}}$,$^{3}$ % TRES
Benjamin V. Rackham$^{\href{https://orcid.org/0000-0002-3627-1676}{\includegraphics[scale=0.5]{plots/orcid.jpg}}}$,$^{6,2}$ % stellar char
Boris S. Safonov$^{\href{https://orcid.org/0000-0003-1713-3208}{\includegraphics[scale=0.5]{plots/orcid.jpg}}}$,$^{10}$ % SAI
Miquel Serra-Ricart$^{\href{https://orcid.org/0000-0002-2394-0711}{\includegraphics[scale=0.5]{plots/orcid.jpg}}}$,$^{4,11,12}$ % TTT
\newauthor
Abderahmane Soubkiou$^{\href{https://orcid.org/0000-0002-0345-2147}{\includegraphics[scale=0.5]{plots/orcid.jpg}}}$,$^{5}$ % stellar char
Georgina Dransfield$^{\href{https://orcid.org/0000-0002-3937-630X}{\includegraphics[scale=0.5]{plots/orcid.jpg}}}$,$^{13,14,1}$ % SPECULOOS
Katharine Hesse$^{\href{https://orcid.org/0000-0002-2135-9018}{\includegraphics[scale=0.5]{plots/orcid.jpg}}}$,$^{2}$ % tfop
Pipa Fernandez de Diego$^{\href{https://orcid.org/0009-0009-6606-6399}{\includegraphics[scale=0.5]{plots/orcid.jpg}}}$,$^{8}$ % WIYN / MEE
\newauthor
Emma Softich$^{\href{https://orcid.org/0000-0002-1420-1837}{\includegraphics[scale=0.5]{plots/orcid.jpg}}}$,$^{7}$ % stellar char / AB
Gregor Srdoc,$^{15}$ % tfop
Mathilde Timmermans$^{\href{https://orcid.org/0009-0008-2214-5039}{\includegraphics[scale=0.5]{plots/orcid.jpg}}}$,$^{1,5}$ % SPECULOOS
Jeroen Audenaert$^{\href{https://orcid.org/0000-0002-4371-3460}{\includegraphics[scale=0.5]{plots/orcid.jpg}}}$,$^{2}$ % TESS NOM author
Roi Alonso$^{\href{https://orcid.org/0000-0001-8462-8126}{\includegraphics[scale=0.5]{plots/orcid.jpg}}}$,$^{4,11}$ % SPECULOOS
\newauthor
Artem~Y.~Burdanov$^{\href{https://orcid.org/0000-0001-9892-2406}{\includegraphics[scale=0.5]{plots/orcid.jpg}}}$,$^{6}$ % SPECULOOS
Yasmin T. Davis$^{\href{https://orcid.org/0009-0000-6625-137X}{\includegraphics[scale=0.5]{plots/orcid.jpg}}}$,$^{1}$ % SPECULOOS
Fatemeh Davoudi$^{\href{https://orcid.org/0000-0002-1787-3444}{\includegraphics[scale=0.5]{plots/orcid.jpg}}}$,$^{5}$ % SPECULOOS
Julien de Wit$^{\href{https://orcid.org/0000-0003-2415-2191}{\includegraphics[scale=0.5]{plots/orcid.jpg}}}$,$^{6}$ % SPECULOOS
Shishir Dholakia$^{\href{https://orcid.org/0000-0001-6263-4437}{\includegraphics[scale=0.5]{plots/orcid.jpg}}}$,$^{16}$ %
\newauthor
Elsa Ducrot$^{\href{https://orcid.org/0000-0002-7008-6888}{\includegraphics[scale=0.5]{plots/orcid.jpg}}}$,$^{17}$ % SPECULOOS
Michaël Gillon$^{\href{https://orcid.org/0000-0003-1462-7739}{\includegraphics[scale=0.5]{plots/orcid.jpg}}}$,$^{5}$ % SPECULOOS
Yilen G\'omez Maqueo Chew$^{\href{https://orcid.org/0000-0002-7486-6726}{\includegraphics[scale=0.5]{plots/orcid.jpg}}}$,$^{18}$ % SPECULOOS
Matthew J. Hooton$^{\href{https://orcid.org/0000-0003-0030-332X}{\includegraphics[scale=0.5]{plots/orcid.jpg}}}$,$^{19}$ % SPECULOOS
\newauthor
Clàudia Janó-Muñoz$^{\href{https://orcid.org/0009-0008-5713-0750}{\includegraphics[scale=0.5]{plots/orcid.jpg}}}$,$^{19}$ % SPECULOOS
Emmanuel Jehin$^{\href{https://orcid.org/0000-0001-8923-488X}{\includegraphics[scale=0.5]{plots/orcid.jpg}}}$,$^{20}$ % SPECULOOS
Akanksha Khandelwal$^{\href{https://orcid.org/ 0000-0003-0335-6435}{\includegraphics[scale=0.5]{plots/orcid.jpg}}}$,$^{18}$ % SPECULOOS
\textcolor{black}{Thomas R. Marriott$^{\href{https://orcid.org/0009-0006-4739-0191}{\includegraphics[scale=0.5]{plots/orcid.jpg}}}$,$^{1}$}
\newauthor
Peter P. Pedersen$^{\href{https://orcid.org/0000-0002-5220-609X}{\includegraphics[scale=0.5]{plots/orcid.jpg}}}$,$^{19,21}$ % SPECULOOS
Manuel Pichardo Marcano$^{\href{https://orcid.org/0000-0003-4436-831X}{\includegraphics[scale=0.5]{plots/orcid.jpg}}}$,$^{18}$ % SPECULOOS
Francisco J. Pozuelos$^{\href{https://orcid.org/0000-0003-1572-7707}{\includegraphics[scale=0.5]{plots/orcid.jpg}}}$,$^{22}$ % SPECULOOS
Rafael Rebolo-López$^{\href{https://orcid.org/0000-0003-3767-7085}{\includegraphics[scale=0.5]{plots/orcid.jpg}}}$,$^{4,11}$ % SPECULOOS
\newauthor
Manu Stalport$^{\href{https://orcid.org/0000-0003-0996-6402}{\includegraphics[scale=0.5]{plots/orcid.jpg}}}$,$^{20,5}$ % SPECULOOS
Alexander Venner$^{\href{https://orcid.org/0000-0002-8400-1646}{\includegraphics[scale=0.5]{plots/orcid.jpg}}}$,$^{23}$ % moral support!
Sel\c{c}uk Yal\c{c}{\i}nkaya$^{\href{https://orcid.org/0000-0002-5224-247X}{\includegraphics[scale=0.5]{plots/orcid.jpg}}}$,$^{24,25,5}$ % SPECULOOS
Francis Zong Lang$^{\href{https://orcid.org/0009-0005-3162-3694}{\includegraphics[scale=0.5]{plots/orcid.jpg}}}$,$^{26}$ % SPECULOOS
\newauthor
Sebastián Zúñiga-Fernández$^{\href{https://orcid.org/0000-0002-9350-830X}{\includegraphics[scale=0.5]{plots/orcid.jpg}}}$,$^{5}$ % SPECULOOS
\\
Affiliations are listed at the end of the paper.
}
\date{Accepted XXX. Received YYY; in original form ZZZ}

\pubyear{\the\year{}}

\begin{document}
\label{firstpage}
\pagerange{\pageref{firstpage}--\pageref{lastpage}}
\maketitle

\begin{abstract}
We present TOI-6981\,b, a cold (\teq, assuming an Earth-like albedo) sub-Neptune (\planetradius) orbiting on a 60.8-day orbit around its M4-type host star (\stellarradius, \stellarmass, \stellarteff), part of the SPECULOOS TEMPOS program.
We validate the planetary nature of TOI-6981\,b using ground-based follow-up photometry, statistical validation and high-resolution spectroscopy.
At such cold temperatures, TOI-6981\,b sits at roughly the location of the water ice line (${\sim}170\,\rm K$). Given its orbital period and high transmission spectroscopy metric (\tsm), it is one of the only planets in the sparsely populated cold planet regime amenable for atmospheric studies with JWST. Therefore, TOI-6981\,b could possibly become a crucial benchmark in cold chemistry studies, tests of Hycean world hypotheses, and drawing parallels to the Solar System's ice giants. Additionally, given its current orbit, it may provide insight into planetary formation and migration at or beyond the water ice line.

\end{abstract}


%%%%%%%%%%%%%%%%%%%%%%%%%%%%%%%%%%%%%%%%%%%%%%%%%%

%%%%%%%%%%%%%%%%% BODY OF PAPER %%%%%%%%%%%%%%%%%%
\section{Introduction}
% \textcolor{red}{Maddy
% % A little something about cold planets, the lack of them, why they're important for formation + migration theories etc, iceline, why M dwarfs are good for this 
% }

While over 6000 exoplanets\footnote{\url{https://exoplanetarchive.ipac.caltech.edu}, March 2026} have been discovered in the last ${\sim}30\,$years, only ${\sim}7\%$ have reported equilibrium temperatures below 400\,K (`temperate' planets, as defined in works such as \citealt{peterson2023}, \citealt{gunther2019}, and formally defined in \citealt{scott2026}). Only ${\sim}1.5\%$ have colder equilibrium temperatures than Earth ($\leq255\,\rm K$).

This apparent scarcity is largely due to observational biases: cold planets orbit at wider separations and therefore have longer orbital periods for which transit and radial velocity (RV) surveys are inherently less sensitive \citep{foremanmackey2016, bryson2020}. 
The geometric transit probability scales inversely with semi-major axis, $a$, as $P_{\rm tr}\sim\frac{R_\star}{a}$, where $R_\star$ is the radius of the host star \citep{borucki_summers1984}, further disfavouring long-period planets. Given the observing strategy of current surveys such as \textit{TESS} \citep[Transiting Exoplanet Survey Satellite;][]{tess}, which typically only monitors a star for ${\sim}27\,$d every ${\sim}2\,$years, the probability of detecting multiple transits of such long-period planets is tragically low.
% This bias is even more pronounced when considering only transiting planets, for which the number of temperate planets drops to 341 ($\sim5.6\%$), and 48 $\leq255\,\rm K$ ($\sim0.8\%$). The geometric transit probability scales inversely with semi-major axis, $a$, as $P_{\rm tr}\sim\frac{R_\star}{a}$, where $R_\star$ is the radius of the host star \citep{borucki_summers1984}. As a result, long-period (and therefore colder) planets are significantly less likely to transit their host stars.
% For example, a planet orbiting a Solar-type star must have an orbital period of at least $\sim 100$ days to reach an equilibrium temperature\footnote{see $^{2}$} of $400\,$K.
% Given the observing strategy of prominent surveys such as \textit{TESS} \citep{tess}, which typically only monitors a star for $\sim27\,$d every $\sim2\,$years, the probability of detecting multiple transits of such long-period planets is tragically low. This is further compounded by their intrinsically small transit probabilities \citep[see e.g. \textit{TESS} planet yields:][]{barclay2018, kunimoto2022}.
M dwarfs, particularly mid-to-late types ($T_{\rm eff}\lesssim 3400\,\rm K$), can substantially minimise this problem. Around these stars, the orbital period required for a planet to reach a given equilibrium temperature is much shorter\footnote{e.g. ${\sim}5\,$days for a typical M4 star at $400\,\rm K$ or $\sim20\,$days at $255\,\rm K$ vs  ${\sim} 100$ days for a Solar-type star at $400\,$K or ${\sim}365\,$days at $255\,\rm K$}, making transits more detectable and enabling feasible follow-up observations with RV measurements and atmospheric characterisation with \textit{JWST}.

% Thankfully, the severity of this problem can be minimised by turning our attention to M dwarfs, particularly those mid- to-late type ($T_{\rm eff}\lesssim 3400\,\rm K$). Around these stars, the orbital period required for a planet to reach a given equilibrium temperature is substantially shorter (e.g. $\sim5\,$days for a typical M4 star at $400\,\rm K$, or $\sim20\,$days at $255\,\rm K$). This significantly increases the likelihood of detecting transits and enables feasible follow-up observations such as RV mass measurements and atmospheric characterisation via transmission spectroscopy with e.g. \textit{JWST}.

Expanding the population of cold, temperate planets is essential for constraining occurrence rates and planetary demographics. Of particular interest are planets located near or beyond the water ice line \citep[$\sim 150-170\,$K;][]{hayashi1981, podolak2004, kennedy2008}, where water transitions from vapour to solid ice.
Planet detections in this cold regime are rare ($\lesssim40$ total and $\lesssim15$ transiting as of March 2026), yet they provide critical insight into planet formation and migration processes; these planets may have formed in-situ through rapid accretion of ice-rich planetesimals \citep{ormel2010,lambrechts2012,venturini2020} and built sufficiently massive cores capable of accreting and retaining a volatile envelope, or alternatively they could have formed even further out in the protoplanetary disk and undergone modest inwards migration through disk-driven processes before the gas dispersed \citep{ida2008a, ida2008b, bitsch2015}. These planets also serve as analogues to the Solar System's giant planets, which formed and remained beyond the Sun's water ice line, providing a rare opportunity to place our Solar System in a broader exoplanetary setting \citep{winn2015}.
% Cold planets occupy a sparsely sampled region of parameter space. Expanding this population is essential for constraining planetary occurrence rates and demographics. As we push to even colder temperatures, we seek to detect and characterise planets located around or beyond the ice line of their host stars. For the purpose of this work, we define the ice line to be at roughly temperatures of $\sim 150-170\,$K, where water transitions from vapour to solid ice (REF).
% Detections of planets sitting $\sim$at or beyond the ice lines of their stars are rare ($\lesssim40$ total and $\lesssim15$ transiting as of March 2026), yet they provide critical insight into planet formation and migration processes; planets in this regime may have formed in-situ through rapid accretion of ice-rich planetesimals and built sufficiently massive cores capable of accreting and retaining a volatile envelope (REFS), or alternatively they could have formed even further out in the protoplanetary disk and undergone modest inwards migration through disk-driven processes before the gas dispersed. 
%Thus, expanding this population of ice-line planets can aid in answering these current questions in planet formation and migration theories.

% Additionally, these cold planets also provide an essential benchmark for comparison with the Solar System's giant planets, Jupiter, Saturn, Uranus and Neptune, which formed beyond the Sun's ice line. Therefore, ice-line exoplanets offer a rare but crucial opportunity to place the full Solar system in a broader exoplanetary context.

In this context, we present TOI-6981\,b, a cold (\teq) sub-Neptune orbiting an M4 host star. Its relatively wide orbit places it well beyond the hot and warm sub-Neptune population commonly observed around M dwarfs, making it a valuable benchmark for testing whether small planets around low-mass stars typically migrate inward or can remain near their formation locations.
As part of the SPECULOOS TEMPOS program \citep{scott2026}, TOI-6981\,b contributes to a growing sample of temperate and cold planets orbiting cool stars. 
Expanding this population is essential for constraining how disk temperature structure, ice-line location and stellar mass influence the formation efficiency, bulk composition and migration histories of small planets.

This paper is organised as follows: in Section~\ref{sec:photometric_obs} we describe the photometric observations of TOI-6981\,b, followed by stellar characterisation of the host star in Section~\ref{sec:stellar_char}. Section~\ref{sec:vetting} outlines the procedures conducted to validate the system, followed by a description of the global analysis of all available data in Section~\ref{sec:analysis}. Finally, we discuss our results in Section~\ref{sec:discussion} and put TOI-6981\,b into context with the current exoplanet population, and describe future prospects for mass and atmospheric characterisation.

% as well as the need for cold planets to exist relatviely far from their star. e.g. solar system planets + state orbital periods.

\section{Photometric Observations} \label{sec:photometric_obs}
A summary of the ground observations described in Sections~\ref{sec:photometric_obs}, ~\ref{sec:stellar_char} and ~\ref{sec:vetting} can be found in Table~\ref{tab:followup_6981}.

\subsection{TESS}
% set up, sectors etc:
TOI-6981 (TIC 443823169, UCAC4 507-044487) has been observed by \textit{TESS} in sectors 7, 34, 44, 45, 46, 61, 71, 72, and 88 with a 120\,s cadence; sectors 61 and 88 with a 20\,s cadence; sectors 61, 71, 72, 88 with a 200\,s cadence; sectors 34, 44, 45, 46 with a 600\,s cadence; and sector 7 with a 1800\,s cadence.
\textcolor{black}{The \textit{TESS} target pixel file for Sector 7 is shown in Fig.~\ref{fig:tpf}, created using \texttt{tpfplotter}\footnote{\url{https://github.com/jlillo/tpfplotter}} \citep{tpfplotter}, where we highlight nearby sources and indicate their magnitude relative to TOI-6981.}

TOI-6981.01 was identified as a planet candidate on UTC 2024 May 30, with a candidate radius of $2.08\pm0.44\,\rm R_\oplus$, a transit depth of $4753\pm396\,\rm ppm$, \textcolor{black}{a maximum orbital period} of $425.45069\pm0.00058\,\rm d$, and a transit duration of $2.95\pm0.47\,\rm hrs$. It was alerted as a TOI from the presence of a transit in Sector 46 followed by one in Sector 61.

% clarify the orbital period:
Upon investigation of the 120\,s cadence \textit{TESS} light curve, an additional transit in Sector 88 was identified via visual inspection that corresponds to the n=7 period alias (i.e., $P=425.45069/7=60.7787\,\rm d$).
Thus the alerted orbital period of the TOI manifested from a duo-transit situation. All other period aliases are ruled out by the 120\,s \textit{TESS} data. Additionally, a KeplerCam observation on UTC 2024 December 08 (see Section~\ref{sec:keplercam}) further rules out all other possible aliases other than the n=7. Therefore, we take this as the true orbital period of the planet candidate. 

% what data we use in this paper:
For the analysis in this paper, we make use of the Presearch Data Conditioning Simple Aperture Photometry \citep[PDCSAP;][]{stumpe2012, smith2012, stumpe2014} 120\,s cadence flux, which is produced by the Science Processing Operations Center \citep[SPOC;][]{jenkins2016} pipeline. 
This flux has been corrected for instrumental systematics and contamination from nearby stars \textcolor{black}{using the TIC-based crowding metric (CROWDSAP=0.90}) offering higher precision than other pipelines. \textcolor{black}{However, we find this contamination correction to be insufficient for TOI-6981 (see Section 5).}
The data are shown in Fig~\ref{fig:tess}, for which we have used \texttt{lightkurve} \citep{lightkurve2018} to access these data from the NASA Mikulski Archive for Space Telescopes (MAST)\footnote{\url{https://mast.stsci.edu}}.

\subsection{Ground-based photometry} %\label{sec:photometric_obs}

\subsubsection{FLWO-1m2 (KeplerCam)} \label{sec:keplercam}

Photometric observations were obtained using the KeplerCam CCD on the 1.2-m telescope at the Fred Lawrence Whipple Observatory (FLWO) at Mount Hopkins, Arizona to observe an ingress of TOI-6981.01 on 2024 December 08. The observation was carried out in the Sloan-$i'$ with an exposure time of 40\,s, and had a FWHM of $2.56\arcsec$ and aperture of $5.4\arcsec$.
KeplerCam is a 4096 x 4096 Fairchild CCD 486 detector which has a field-of-view of $23\farcm1\times23\farcm1$ and an image scale of 0.672$"/$pix when binned by 2.
Transit observations were scheduled using the \textit{TESS} Transit Finder, a customized version of the \textsc{TAPIR} package \citep{Jensen2013}. 
Data were reduced using standard IDL routines and aperture photometry was performed using {\tt AstroImageJ} software \citep{Collins_2017}.

\subsubsection{SPECULOOS-North-1m0 (Artemis)}
A full transit of TOI-6981.01 was observed with the SPECULOOS \citep[Search for Planets EClipsing ULtra-cOOl Stars;][]{Sebastian_2021AA,Burdanov2022}-North Artemis telescope located at Teide Observatory on UTC 2026 February 6. SPECULOOS-North/Artemis is equipped with a 2k$\times$2k Andor iKon-L camera. It has FOV of $12\,\arcmin\times12\,\arcmin$ and a pixel scale of 0.35 arcsec. 
The transit was observed in the \textit{Sloan-r'} filter with an exposure time of 18\,s. The observation had a FWHM of $2.6\arcsec$ and an aperture of $3.5\arcsec$.
The data was reduced using the {\tt prose}\footnote{{\tt Prose:} \url{https://github.com/lgrcia/prose}} pipeline \citep{prose_2022}.

\subsubsection{LCOGT-1m0}
% \textcolor{red}{khlaid / Karen?}

A full transit of TOI-6981.01 was observed with the LCOGT \citep[Las Cumbres Observatory Global Telescope;][]{Brown_2013} 1m0 telescope located at Teide Observatory on UTC 2026 February 6. The telescope is equipped with 4096$\times$4096 SINISTRO camera, having a FOV of 26\,arcmin$\times$26\,arcmin and a pixel scale of 0.389 arcsec. 
The transit was observed in the \textit{Sloan-i'} filter with an exposure time of 40\,s, as well as a FWHM of $3.3\arcsec$ and aperture of $4.7\arcsec$.
The data was reduced using {\tt BANZAI} pipeline \citep{McCully_2018SPIE10707E}, and photometric extraction was performed in 4.7$\arcsec$ uncontaminated aperture using {\tt AstroImageJ} software.

\subsubsection{TTT-2m0}
% \textcolor{red}{khalid} + Miquel

A full transit of TOI-6981.01 was observed with the Two-meter Twin Telescope (TTT\footnote{\url{https://ttt.iac.es/}}), located at the Teide Observatory, Canary Islands, Spain (28$^\circ$18$'$04$''$ N, 16$^\circ$30$'$38$''$ W, alt. 2362 m). The facility consists of two identical 2.0-m f/6 Ritchey-Chrétien telescopes (TTT3/4), with 
% and two smaller replicas of 0.80-m f/6.85 (TTT1/2), all mounted on Alt-Az systems and equipped with multiple instruments at the Nasmyth foci. The 0.80-m telescopes have been operational since 2023 (MPC codes Y65, Y66), while 
the first 2.0-m unit (TTT3, Y68) being under commission since early 2025. All telescopes are equipped with FERVOR (\textit{Fast Embedded-sCMOS Robotic Visible Observatory for Rapid transients}), an optical imager based on sCMOS sensors (Sony IMX455 and IMX411; \citealt{Alarcon2023}). The medium-format version, installed on TTT3, provides $10.2'\times6.8'$ with a plate scale of $0.19''~{\rm px^{-1}}$ when operated with $3\times3$ binning. These detectors have negligible dark current, virtually no readout overhead, a minimum exposure time of 1\,s, and very low readout noise ($4.7~e^-$). The transit was observed on UTC 2026 February 6 in the \textit{Pan-STARRS $\rm z_s$} filter with an exposure time of 20\,s. It had a FWHM of $2.45\arcsec$ and an aperture of $4.2\arcsec$.
The data reduction was done in the standard way, and photometric extraction was performed in 4.2$\arcsec$ uncontaminated aperture using {\tt AstroImageJ} software.

\section{Stellar Characterisation}  \label{sec:stellar_char}

% ==========================================
% stellar params table:
\begin{table}
\centering
\caption{Stellar parameters adopted for this work. 
}
\begin{tabular}{@{}lp{40mm}p{20mm}@{}}
\toprule
{\bf Star} & {\bf TOI-6981} & \\
\toprule
% {\bf Designations} & \multicolumn{1}{p{40mm}}{TIC 443823169, 2MASS J08041456+1116015, Gaia DR2 3147904852637255424, UCAC4 507-044487} & \\ \midrule
{\bf Designations} & TIC 443823169, 2MASS J08041456+1116015, Gaia DR2 3147904852637255424, UCAC4 507-044487 & \\ \midrule
{\bf Parameter} & {\bf Value}  & {\bf Source} \\ \midrule
T mag      &  $11.663\pm0.007$     & \cite{TICv8} \\
B mag      &   $15.917\pm0.033$     & \cite{ucac4} \\
V mag      &     $14.338\pm0.038$      & \cite{ucac4} \\
G mag      &      $12.9676\pm0.0011$     & \cite{gaiaDR3cat} \\
J mag      &    $9.97\pm0.023$         & \cite{2masscat} \\
H mag      &      $9.413\pm0.03$      & \cite{2masscat} \\
K mag      &     $9.151\pm0.02$       & \cite{2masscat} \\
W1 mag     &   $8.651\pm0.029$   & \cite{wisecat} \\
W2 mag     &    $8.529$      & \cite{wisecat} \\
W3 mag     &    $8.994\pm0.022$     & \cite{wisecat} \\
W4 mag     &    $8.831\pm0.02$   & \cite{wisecat} \\
Distance   &  $25.47\pm0.05$\,pc     & \cite{BJdist} \\
Parallax & $39.48\pm0.03\rm\,mas$ & \cite{gaiaDR3cat} \\
$\alpha$    &    08:04:14.39        & \cite{gaiaDR3cat} \\
$\delta$     &  +11:16:00.97        & \cite{gaiaDR3cat} \\
$\mu_{\alpha}$   &    	$\rm -166.176\,mas\,yr^{-1}$       & \cite{gaiaDR3cat} \\
$\mu_{\delta}$   &   $\rm -39.1326\,mas\,yr^{-1}$        & \cite{gaiaDR3cat} \\
SpT      &  M4              & This work (opt. spec.) \\
     &  M3.5$\pm$0.5                 & This work (NIR spec.) \\
$R_{\star}$  & $0.304\pm 0.012\,\mathrm{R}_{\odot}$  & This work (SED)             \\

$M_{\star}$  & $0.296\pm 0.026\,\mathrm{M_{\odot}}$  & This work (SED)   \\

${T_{\rm eff}}$ & $3215\pm 58$\,K         & This work (SED)              \\
 
$\log g_\star$   &   $4.942\pm 0.035$        & This work (SED)                 \\

$\rm [Fe/H]$     & $+0.06\pm 0.11$ dex          & This work (SED)                 \\
                & $+0.06\pm 0.20$ dex      & This work (opt. spec.)                 \\
                & $+0.05\pm 0.12$ dex              & This work (NIR spec.)                 \\    
Age & $\leq1.5\,\rm Gyr$ & This work (H$\alpha$ emission) \\
         
\bottomrule
\end{tabular}
\label{tab:starpar}
\end{table}

% ==========================================

\subsection{Spectroscopic characterisation}

\subsubsection{IRTF/SpeX} %BVR

We observed TOI-6981 with the SpeX spectrograph \citep{Rayner2003} on the 3.2-m NASA Infrared Telescope Facility (IRTF) on 12 December 2024 (UT) under clear conditions with $1\farcs0$ seeing.
We used the short-wavelength cross-dispersed (SXD) mode and the $0\farcs3 \times 15''$ slit ($R{\sim}2000$, 0.80–2.42\,$\mu$m) aligned to the parallactic angle to gather six 60\,s exposures at an airmass of 1.0, nodding in an ABBA pattern.
Science observations were followed by a standard set of SXD flat field and arc lamp calibrations and six 20\,s exposures of the A0\,V standard HD\,58383 ($V{=}7.4$) at a similar airmass.
The data were reduced with the Spextool v4.1 pipeline \citep{Cushing2004}, following the standard approach \citep{Barkaoui2024, Barkaoui2025, Ghachoui2024}.
The final spectrum has a median SNR per resolution element of 127.

\begin{figure}
    \centering
    \includegraphics[width=\linewidth]{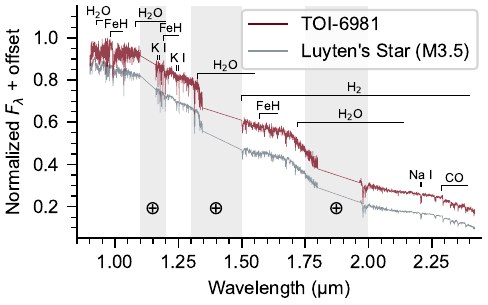}
    \caption{
    SpeX SXD spectrum of TOI-6981.
    The target spectrum (red) is shown alongside that of the M3.5V standard Luyten’s Star (grey), offset vertically for clarity.
    Regions of strong telluric absorption are shaded, and strong M-dwarf spectral features are indicated.
    }
    \label{fig:spex}
\end{figure}

The SXD spectrum of TOI-6981 is shown in Fig.\,\ref{fig:spex}.
Using the SpeX Prism Library Analysis Toolkit \citep[SPLAT, ][]{splat} to compare the spectrum to single-star standards in the IRTF Spectral Library \citep{Cushing2005, Rayner2009}, we find the closest match to the M3.5\,V standard Luyten’s Star and adopt a spectral type of M3.5\,$\pm$\,0.5.
From the H$_2$O--K2 index \citep{Rojas-Ayala2012} and $K$-band Na\,\textsc{i} and Ca\,\textsc{i} line strengths, we estimate $\mathrm{[Fe/H]} = +0.05 \pm 0.12$ using the \citet{Mann2013} relation.

\subsubsection{Shane/Kast}

TOI-6981 was observed with the Kast double spectrograph \citep{kastspectrograph} on the 3-m Shane telescope at Lick Observatory on UTC 2024 December 5 in clear and windy conditions with $\sim$2$\arcsec$ variable seeing. We used the 2$\farcs$5 slit aligned to the parallactic angle to obtain blue and red optical spectra split at 5700\,{\AA} by the d57 dichroic, and dispersed by the 600/4310 grism and 600/7500 grating, resulting in spectral resolutions of $\lambda/\Delta\lambda$ $\approx750$ and $\approx1000$, respectively. We obtained a single 500\,s exposure in the blue channel and two 250\,s exposures in the red channel at an average airmass of 1.26. The G2\,V star HD 91950 ($V=8.6$) was observed immediately before the TOI-6981 observations at a similar airmass for telluric absorption calibration, and the spectrophotometric calibrator Hiltner~600 \citep{1992PASP..104..533H,1994PASP..106..566H} was observed on the same night for flux calibration.  HeHgCd and HeNeArHg arc lamps were used to wavelength calibrate the blue and red data, respectively, and white light flat-field lamp exposures were used to correct for pixel response variation. Data were reduced using the \texttt{kastredux} code\footnote{\url{https://github.com/aburgasser/kastredux}} \citep{kastredux} using standard settings. The resulting spectra have median SNRs of 50 at 5425\,{\AA} and 160 at 7350\,{\AA}.

\begin{figure}
    \centering
    \includegraphics[width=\linewidth]{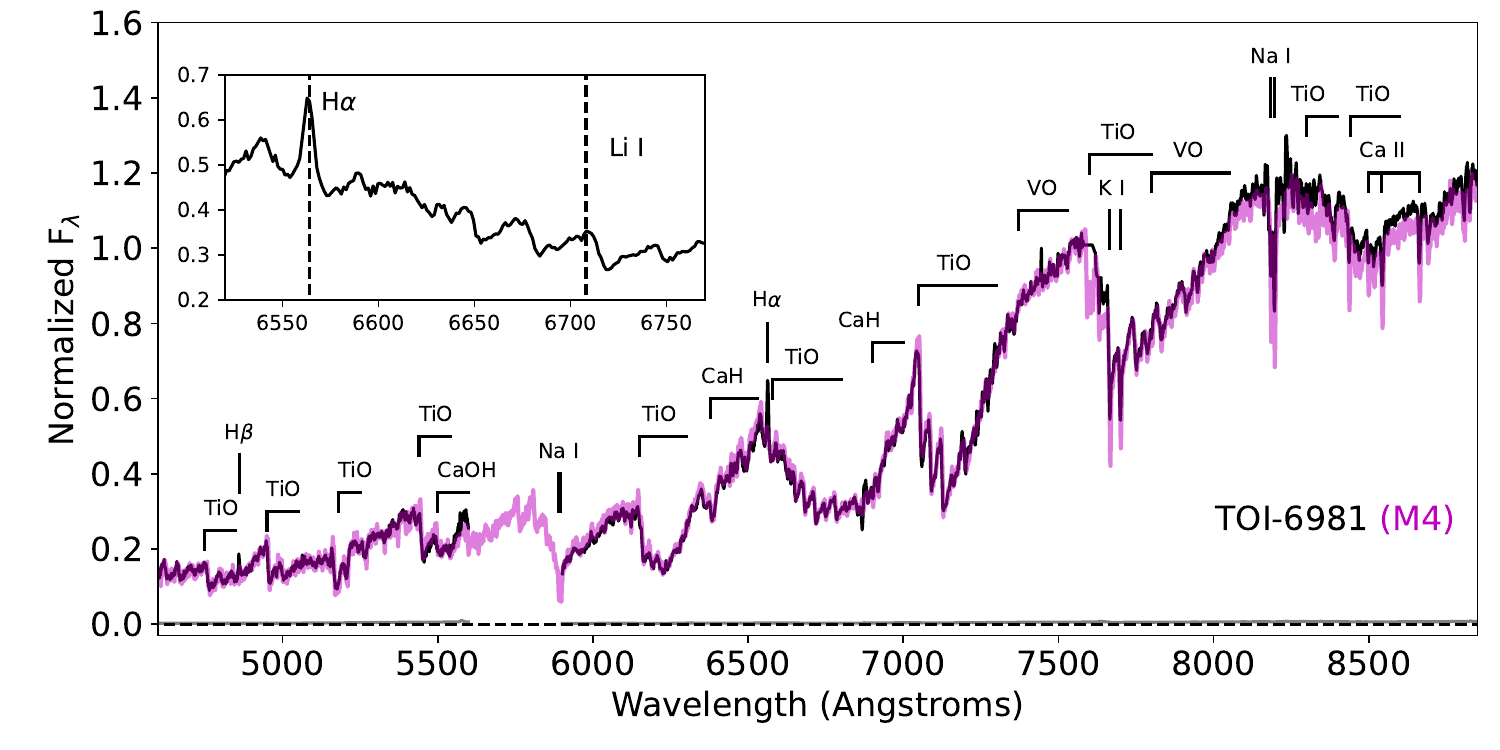}
\caption{Shane/Kast optical spectrum of TOI-6981 (black line) compared to its closest-match M4 SDSS spectral template from \citet[magenta line]{2007AJ....133..531B}. Primary atomic and molecular spectral features are labelled, including emission lines from H$\beta$ (4861\,{\AA}) and H$\alpha$ (6563\,{\AA}). 
%The gap in data between 5600~$\AA$ and 5900~$\AA$ corresponds to the gap between the blue and red channels of Kast.
The inset box shows the 6520--6770\,{\AA} region encompassing the H$\alpha$ emission line and absent Li\,\textsc{i} (6708\,{\AA}) absorption.}
    \label{fig:kast}
\end{figure}

The reduced optical spectrum of TOI-6981 is shown in Fig.\,\ref{fig:kast}, compared to its best-match M4 dwarf spectral template from 
\citet{2007AJ....133..531B}. This template type splits prior spectral classifications of M5 from \citet{2014AJ....147...20N} based on near-infrared observations and M3 from \citet{2023ApJS..264...17Z} based on optical LAMOST data, and is in good agreement with our near-infrared SpeX-based spectral classification.
Classifications based on optical spectral indices defined by
\citet{1995AJ....110.1838R,1997AJ....113..806G,1999AJ....118.2466M,2003AJ....125.1598L}; and \citet{2007MNRAS.381.1067R} similarly span M3.5--M4.
We measured a zeta metallicity index of $\zeta=1.039\pm0.002$  \citep{2013AJ....145..102L}, corresponding to [Fe/H] = $+0.06\pm0.20$ \citep{Mann2013}, again consistent with our SpeX measurements
and with prior estimates based on infrared spectroscopy
([Fe/H] = +0.13$\pm$0.13; \citealt{2014AJ....147...20N})
and optical-infrared photometry
([Fe/H] = +0.19$\pm$0.11; \citealt{2016ApJ...818..153D})
We detected clear hydrogen emission lines at 4861\,\AA\ (H$\beta$) and 6563\,\AA\ (H$\alpha$) 
with equivalent widths of $-1.76\pm0.16$\,{\AA} and $-1.58\pm0.06$\,{\AA}, respectively,
confirming prior reports that TOI-6981 is an active ``flare star'' \citep{2023ApJS..264...17Z}.
The H$\alpha$ emission implies a relative H$\alpha$ luminosity of $\log{\left(L_{{\rm H}\alpha}/L_{\rm bol}\right)} $=$ -4.25\pm0.07$ using the $\chi$ factor/spectral type relation of \citet{2014ApJ...795..161D}, and indicates an activity age $\lesssim$3.5--5\,Gyr \citep{2008AJ....135..785W}.
We see no evidence of Li\,\textsc{i} at 6708\,{\AA} Li\,\textsc{i}, ruling out a substellar object with an age less than ${\sim}30$\,Myr.

\subsection{Spectral energy distribution} \label{sec:sed}

\begin{figure}
	\centering
	\includegraphics[width=0.5\textwidth]{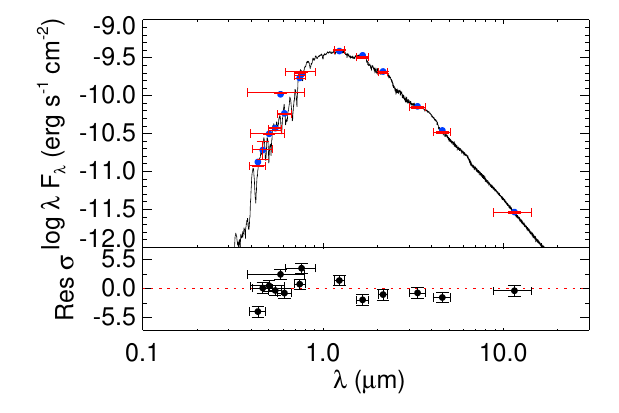}
	\caption{Spectral energy distribution (SED) fit of flux as a function of wavelength for TOI-6981. Blue points are the best-fit values, and red points are the corresponding model values and errors.}
	\label{fig:sed}
\end{figure}

To derive the stellar parameters, we used the publicly available exoplanet fitting suite \texttt{EXOFASTv2} \citep{Eastman2017, Eastman2019}, which employs a differential evolution Markov chain Monte Carlo (DE-MCMC) method to jointly fit stellar and exoplanetary parameters. We simultaneously fit the stellar spectral energy distribution (SED) while incorporating the MESA Isochrones and Stellar Tracks (MIST; \citealt{Paxton2015}). For the SED fit, we used broadband apparent magnitudes from 2MASS ($J$, $H$, and $K_s$), WISE ($W1$, $W2$, and $W3$), \textit{Gaia} EDR3 ($G$, $G_{\rm BP}$, and $G_{\rm RP}$), and APASS ($B$, $V$, $g$, $r$, and $i$) (see Fig.~\ref{fig:sed}). These photometric data were obtained using \texttt{MKSED}, a utility distributed with \texttt{EXOFASTv2} that automatically queries standard photometric catalogs. Gaussian priors were imposed on the \textit{Gaia} DR3 parallax \citep{Gaia2020}, after adding the offset calculated following the prescription of \citet{Lindegren2021}. An upper limit on the $V$-band extinction, $A_V < 0.077$, was adopted based on reddening maps \citep{Schlegel1998, Schlafly2011}, with the fit returning  $A_V = 0.039 \pm 0.026$ mag. The resulting stellar parameters are reported in Table~\ref{tab:starpar}.

\subsection{Empirical relations} \label{sec:emp_rels}  %BVR

Using empirical relations from \citet{Mann2015, Mann2019}, we derive updated stellar parameters for TOI-6981. 
From its apparent magnitudes and \textit{Gaia} DR3 parallax, we estimate a stellar mass of $0.2682 \pm 0.0061\,M_\odot$ and a radius of $0.2955 \pm 0.0084\,R_\odot$, corresponding to a surface gravity of $\log(g) = 4.925 \pm 0.027$\,dex. 
The \textit{Gaia} DR3 BP--RP color implies an effective temperature of $3356 \pm 77$\,K. 
Combining the radius and effective temperature via the Stefan--Boltzmann law yields a stellar luminosity of $0.0100 \pm 0.0011\,L_\odot$.

\subsection{\textcolor{black}{Choice of stellar parameters for this work}}

We obtain stellar parameters from both SED fitting (Section~\ref{sec:sed}) and empirical relations (Section~\ref{sec:emp_rels}). For the analysis presented in this work, we choose to adopt the stellar parameters obtained from the SED fit (see Table~\ref{tab:starpar}). 
The use of a wide range of broad-band photometry and the allowance of $T_{\rm eff}$ and metallicity to be free parameters means that the resulting parameters from SED fitting should be more robust, with uncertainties that are more conservative. Conversely, the parameters derived from empirical relations rely on \textit{Gaia} DR3 BP-RP colour for estimations of the stellar $T_{\rm eff}$, something which is not entirely reliable for cool, low-mass stars \citep{jordi2010}. Additionally, it is sensitive to metallicity which is a parameter difficult to robustly determine for stars of this spectral type \citep{lindgren2017, passegger2022}.

\section{Vetting} \label{sec:vetting}

\subsection{High resolution imaging} \label{sec:highres_imaging}

\subsubsection{Zorro} \label{sec:zorro}
%\textcolor{red}{Steve Howell}
\textcolor{black}{If a host star} has a spatially close companion, that companion (bound or line of sight) can create a false-positive transit signal if it is, for example, an eclipsing binary (EB). Alternatively, ``third-light” flux from the close companion star can lead to an underestimated planetary radius and incorrect physical parameters if not accounted for in the transit model \citep{ciardi2015, furlanhowell2017, furlanhowell2020}. 
Thus, to search for close-in bound companions unresolved in \textit{TESS} or other ground-based follow-up observations, we obtained high-resolution imaging speckle observations of TOI-6981. TOI-6981 was observed on UT 2025 February 16 using the `Alopeke speckle instrument on the Gemini North 8-m telescope \citep{scottN2021}. `Alopeke provides simultaneous speckle imaging in two bands (562\,nm and 832\,nm) with output data products including a reconstructed image with robust contrast limits on companion detections. Six sets of $1000 \times 0.06$ sec exposures were collected and subjected to Fourier analysis in the standard reduction pipeline \citep[see][]{howell2011}. Figure~\ref{fig:zorro} shows our final 5$\sigma$ magnitude contrast curves and the 832\,nm reconstructed speckle image. We find that TOI-6981 is a single star with no companion brighter than 5 to nearly 9 magnitudes below that of the target star from the diffraction limit (20 mas) out to $1\farcs2$. At the distance of TOI-6981 (d=25.5\,pc) these angular limits correspond to spatial limits of 0.5 to 31\,au.

\subsubsection{WIYN/NESSI}
We observed TOI-6981 on UT 2025 February 4 using the NN-EXPLORE Exoplanet Stellar Speckle Imager \citep[NESSI;][]{scott2018} at the WIYN 3.5-m telescope on Kitt Peak. The data consist of 9000 40\,ms frames in two filters with central wavelengths $\lambda_c = 562$ and 832\,nm.  A similar set of 1000 speckle frames were taken of a nearby, single star in order to calibrate the PSF of the TOI-6981 data. These speckle data were reduced using the pipeline described by \citet{howell2011}. Among the pipeline products is a reconstructed image of the field surrounding the target in each filter.  We obtained a contrast curve (see Figure~\ref{fig:nessi}) from each reconstructed image by measuring fluctuations in the background level as a function of separation from TOI-6981 out to a maximum separation of $1.2\arcsec$. These contrast curves establish the upper limits on the relative brightness of any undetected point source in close proximity to the target star. No companion sources were detected for TOI-6981 in the NESSI data.

\subsubsection{SAI}

TOI-6981 was observed on 2024 November 14 with the speckle polarimeter on the 2.5-m telescope at the Caucasian Observatory of Sternberg Astronomical Institute (SAI) of Lomonosov Moscow State University \citep{Strakhov2023}. A low noise CMOS Hamamatsu ORCA-quest was used as a detector. The atmospheric dispersion compensator was active, which allowed using the $I_\mathrm{c}$ band. The angular resolution is $0.083^{\prime\prime}$. Long exposure seeing, as estimated from the data, was good, $0.73^{\prime\prime}$. No companions were detected. The detection limits at distances $0.25$ and $1.0^{\prime\prime}$ are $\Delta I_\mathrm{c}=4.5^m$ and $5.6^m$, see Fig.~\ref{fig:SAI_speckle}.

\subsection{TRES spectroscopy} \label{sec:TRES}
Three observations of TOI-6981 each with exposure times of 3600\,s were collected with the Tillinghast Reflector Echelle Spectrograph (TRES) on the 1.5-m Tillinghast Reflector at the Fred Lawrence Whipple Observatory, which observes at $R = 44,000$ over a wavelength range of 390--910\,nm \citep{Szentgyorgyi2007}. We reduced these observations using a pipeline tailored for mid-to-late M dwarfs, as described in \citet{Pass2023}.

We do not identify any statistically significant RV variation between the observations taken on 26 October 2025, 25 November 2025, and 22 February 2026. The individual RVs are 38.368\,kms$^{-1}$, 38.371\,kms$^{-1}$, and 38.383\,kms$^{-1}$, with per-observation relative RV uncertainties of 20\,ms$^{-1}$. Our systemic RV for TOI-6981 is \hbox{38.37 $\pm$ 0.50\, km\,s$^{-1}$,} which accounts for the uncertainty in the absolute RV of the Barnard's Star template that was used to set the RV zero point \citep{Winters2026}. 
\textcolor{black}{These RVs cover the phases $\sim0.81\,$, $\sim0.30\,$ and $\sim0.76\,$ assuming the planet period of $\sim60.77\,$d, as shown in Fig.~\ref{fig:tres}}.

We do not observe rotational broadening in the TRES spectra, placing an upper limit of $v\sin i$ < 3.4\,kms$^{-1}$. This null detection is expected given the known rotation period of this star (43 d; \citealt{Newton2016}). We do, however, observe H\textalpha\ emission, with an equivalent width of $-2.3\pm0.3$\,\AA, measured following the methods of \citet{Medina2020} and \citet{Pass2023b}; the quoted error reflects the variability in the H\textalpha\ feature over the three observations. Using the mass-dependent activity lifetime equation from  \citet{Pass2024} that is calibrated using the volume-complete sample of single mid-to-late M dwarfs within 15\,pc \citep{Winters2026}, the presence of H\textalpha\ emission implies an age $<$1.5\,Gyr. These findings are consistent with the presence of H\textalpha\ emission in the lower-resolution Kast spectra and place a tighter upper limit on the age. This relatively young age is also consistent with the galactic kinematics of this star, as the UVW space motion of TOI-6981 is similar to that of other active mid-to-late M dwarfs when compared to the sample from \citet{Pass2024}.

We do not see evidence for double lines in the spectra from visual inspection of cross correlation and least-squares deconvolution \citep{Donati1997} plots. Nonetheless, we investigate the spectra with the \texttt{TODCOR} algorithm \citep{Zucker1994} to search for a faint second set of lines, following the implementation for TRES described in \citet{Winters2020} and available in the \texttt{tres-tools}\footnote{\href{https://github.com/mdwarfgeek/tres-tools}{https://github.com/mdwarfgeek/tres-tools}} package. Companions with magnitude differences $\leq$ 2.5 (flux ratios $\leq$10) are reliably detected using this technique \citep{Winters2018}, but we do not identify any such companions for TOI-6981. We cannot rule out cases where both sets of lines are nearly overlapping across the four months of observation (i.e., long-period binaries) or cases where the secondary is substantially fainter than a flux ratio of 10; however, the former case is disfavoured by the lack of companions in high-resolution imaging, and the presence of a close-in but faint secondary is disfavoured by the lack of RV variation across our three measurements.

\subsubsection{Detection limit with \texttt{kima}}
\begin{figure}
    \centering
    \includegraphics[width=1\linewidth]{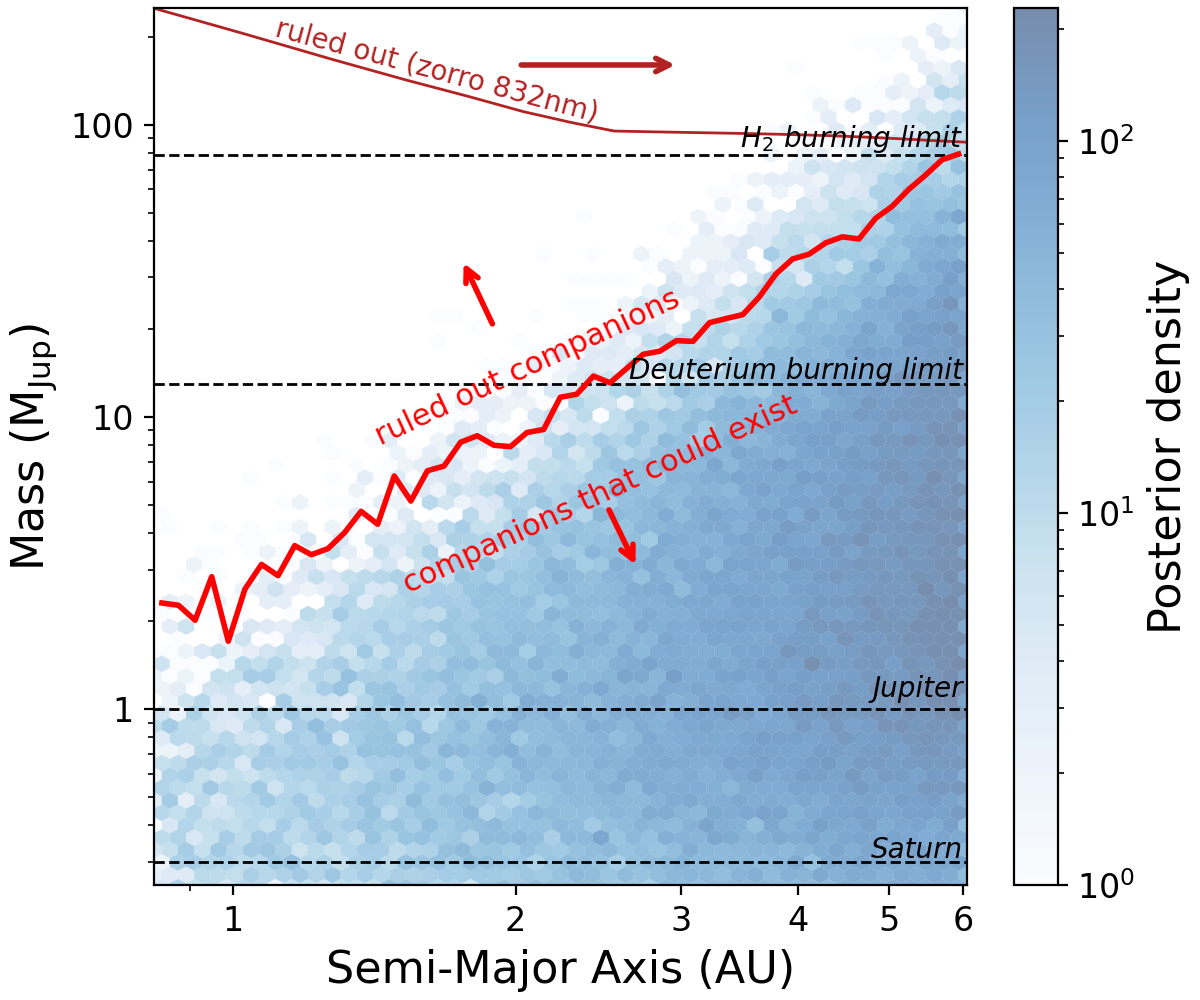}
    \caption{Detection sensitivity limit for TOI-6981. The blues show the posterior density of a \texttt{kima} run forced to fit a singular Keplerian signal to the three TRES RV measurements. The thicker, bright red line indicates the 99th\% upper limit in mass. The signals expected from Saturn and Jupiter, as well as the deuterium ($\sim13\,\rm M_{Jup}$) and hydrogen ($\sim80\,\rm M_{Jup}$) burning limits are shown with black dashed lines.
    We include as reference the 832\,nm Zorro sensitivity curve (the thinner, darker line, also see Section~\ref{sec:zorro}), where we use 1\,Gyr isochrones to approximately convert the magnitude difference to a stellar mass \citep[][using an average of the the CFHT I- and R-bands in order to interpolate from TOI-6981's magnitude as measured in the i-band]{baraffe2015}. 
    We indicate the regions of parameter space that are ruled out by all of these observations, 
    highlighting that up to a separation of 6\,AU, 
    no stellar companions are compatible.
    }
    \label{fig:sensitivity_limit}
\end{figure}

%\textcolor{red}{Aleyna A.}
 We compute a detection sensitivity limit using the three TRES RV measurements to constrain potential other signals in this system. This is calculated following the same method as outlined in \citet{Standing2022, Standing2023, Standing2026} by using \texttt{kima} \citep{kima}, which implements a diffusive nested sampler to explore the parameter space and fit Keplerian curves to RV measurements. Here, the number of Keplerian signals that \texttt{kima} fits is fixed to one, forcing the nested sampler to explore solutions compatible with one signal despite none formally being detected. This analysis results in a posterior distribution of all Keplerian signals that are compatible with the RV measurements but remain undetected. For each signal in the resulting posterior distribution the mass and semi-major axis are calculated. Figure~\ref{fig:sensitivity_limit} shows a density plot of semi-major axis versus companion mass calculated for each signal in the posterior distribution along with a 99th\% upper limit in mass. Sensitivity is reached for Jupiter mass planets at short orbital periods and brown dwarfs at longer orbital periods. 
 % It is also possible to exclude the presence of a companion with a mass larger than a brown dwarf.
 We also include the Zorro sensitivity curve in the 832\,nm band (see Section~\ref{sec:zorro} and Fig~\ref{fig:zorro}) where we use the \citet{baraffe2015} 1\,Gyr isochrones\footnote{1\, Gyr CFHT photometric bands isochrones obtained from \url{https://perso.ens-lyon.fr/isabelle.baraffe/BHAC15dir/BHAC15_iso.CFHT}. The use of this isochrone is motivated by the upper limit constraint on the age of TOI-6981 being $\leq1.5\,\rm Gyr$.} to convert from the Zorro $\Delta m$ into an approximate companion mass.
To do this, we take an average of the CFHT I-band and R-band magnitude values at a given stellar mass to get an approximate \textit{i}-band magnitude ($\sim$ Zorro) and interpolate. Taking the stellar magnitude of TOI-6981 in \textit{i}-band ($i=12.453\pm0.001$), we calculate the $\Delta m$ and find the corresponding stellar masses from the interpolation function.
 We show how a combination of the TRES RV data can rule out larger regions of parameter space than just the Zorro data (and the other high-resolution imaging observations described in Section~\ref{sec:highres_imaging}), and conclude that up to 6\,au, no stellar companions (i.e., more massive than a brown dwarf) are compatible. 

\subsection{Archival Imaging}
%\textcolor{red}{Khalid}

We explored the archival data for TOI-6981 for excluding any possible background objects that could be blended with our target at its current position. 
The target has a proper motion of $\sim 170\,\rm mas\,yr^{-1}$. 
We used the data from POSS-I \citep{1963POSS-I} in 1951 (in the red filter), and POSS-II \citep{1996DSS_POSS-II} in 1988 and 1999 in the red and near-infrared  filters, respectively. We obtained new data from TTT-2.0m in 2026 in the \textit{Pan-STARRS $\rm z_s$} filter. TOI-6981 is shifted by $\sim 12.7\arcsec$ from 1951 to 2026. No background objects are blending within the current position of TOI-6981, as shown in Figure\,\ref{fig:archival}.

\subsection{Statistical Validation}
Following the processes outlined in similar works such as \citet{cacciapuoti2022, harris2023, chew2026}, we use the \texttt{triceratops} \citep{giacalone2020, giacalone2021} to validate the planetary nature of TOI-6981.01 by assessing the likelihood that the transit event is produced by something other than a planet transiting the target star. To be considering validated, we require the false positive probability $\mathrm{FPP}\leq0.015$ and the nearby false positive probability $\mathrm{NFPP}\leq10^{-3}$.
We use the \textit{i'} transit from LCOGT-1m0, and find a $\mathrm{FPP}$ and a nearby false positive probability ($\mathrm{NFPP}$) to both be $\leq 4.2\times10^{-5}$, although we note that since a transit TOI-6981\,b was detected from the ground, the NFPP can already be considered 0.
Thus, TOI-6981.01 (hereafter TOI-6981\,b) is statistically validated under these criteria.

\section{Global Analysis} \label{sec:analysis}
\begin{figure}
    \centering
    \includegraphics[width=1\linewidth]{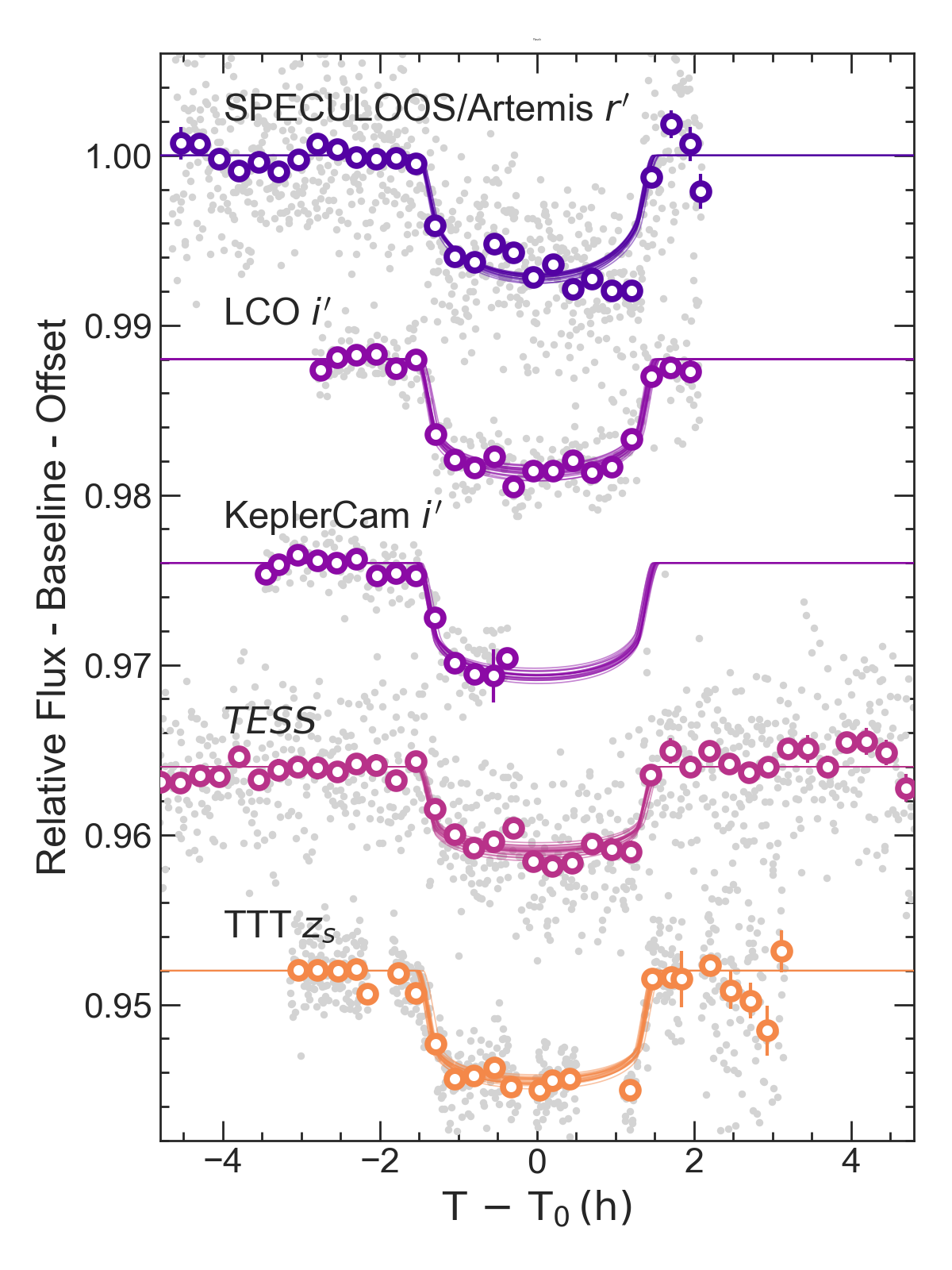}
    \caption{\textcolor{black}{Phase-folded fitted transits of TOI-6981\,b from \textit{TESS} 120\,s cadence and ground-based observations. Raw flux points are in grey and binned (15 mins) flux points are shown by the white circles, colour-outlined to indicate the respective observation filters. 
    Transit models have been corrected by subtracting the baseline and are plotted relatively offset from each other for visualisation purposes.
    The model lines consist of 20 random draws from the posterior transit model.}
    }
    \label{fig:fitted_transits}
\end{figure}

\color{black}
\begin{figure}
    \centering
    \includegraphics[width=1\linewidth]{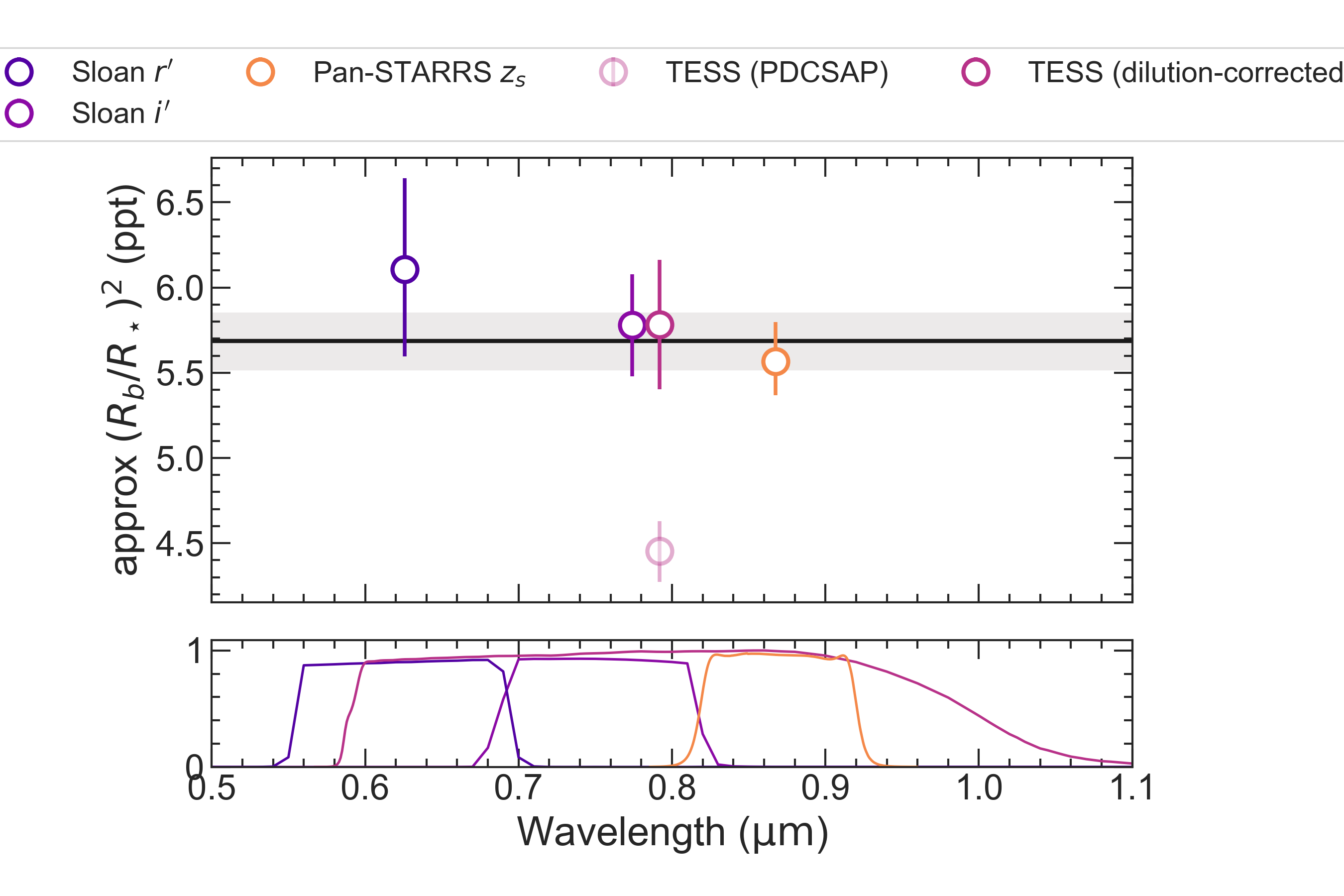}
    \caption{
    \color{black}
    Chromaticity check of TOI-6981\,b. Depths are obtained from the chromatic (i.e., free dilution) eccentric fit and are corrected for limb darkening; the two Sloan-\textit{i'} depths (LCO and KeplerCam) are combined by inverse-variance weighting.
    The black line and grey band show the inverse-variance weighted mean of the ground-based depths and its $1\,\sigma$ uncertainty.
    The faded \textit{TESS} point shows the PDCSAP depth, which lies $5.1\sigma$ below this mean (and we note that all three \textit{TESS} transits share a consistent depth). The second \textit{TESS} point is corrected for the \textit{TESS} dilution from the final fit ($D_{\rm TESS} = 0.23\pm0.04$; Section~\ref{sec:analysis}). As this correction is calibrated against the ground-based depths, the corrected point is not independent and is excluded from the weighted mean.
    }
    \label{fig:chromaticity}
\end{figure}

\color{black}
Following the methods outlined in e.g., \citet{dransfield2024} and \citet{scott2026}, we use \texttt{allesfitter} \citep{allesfitter-code, allesfitter-paper}, to fit all transits simultaneously. For the reasons laid out in these works, we opt for the nested sampling mode in \texttt{allesfitter} for our fit, \textcolor{black}{however we summarise again here for clarity.
\texttt{allesfitter} offers both a nested sampling mode \citep[via \texttt{dynesty};][]{speagle2020} and an MCMC mode  \citep[via \texttt{EMCEE};][]{foremanmackey2013}. Since the nested sampling approach allows for direct calculation of the Bayesian evidence \citep[via the Bayes factor;][]{BayesFactor}
, this allows us to compare different models (for example, circular vs eccentric), and determine which, if any, is statistically favoured.}

We obtain limb darkening priors from 
\textsc{pyltdk} \citep[][reparameterised as in \citealt{kipping2013}]{pyldtk}, and all other priors from the candidate planetary values and stellar parameters outlined in Section~\ref{sec:photometric_obs} and Table~\ref{tab:starpar}. Priors are shown in Table~\ref{tab:priors}.
We note that we use an informative prior on the limb darkening since the data are not of good enough quality to properly inform the fit. For these parameters, the fit returns the prior.
\textcolor{black}{For the \textit{TESS} PDCSAP photometry, we fit for an additional dilution, $D_{\rm TESS}$ with a uniform prior $\mathcal{U}(-1,1)$, while fixing the dilution to zero for all ground-based instruments, which were extracted in uncontaminated apertures (see Section~\ref{sec:photometric_obs}). The motivation for including this dilution is described below in Section~\ref{sec:chrom_check}.}

To account for out-of-transit flux, we use a hybrid spline to model any variation. Since we take a window of 0.5 days centred around the transit mid-point, we negate the need for more complex models (e.g., Gaussian Processes) to account for additional activity seen in the light curve (e.g., the flares in \textit{TESS}).

We fit for two models, circular and free eccentricity, and calculate the log marginal likelihood difference, $\Delta \log Z$. 
We find that at present, the \textcolor{black}{eccentric} model is favoured with a \textcolor{black}{$\Delta \log Z=3.1$}, thus we adopt this fit for the analysis in this work.
%We find that neither model is favoured, with a $\Delta \log Z=-1.02$. 
% However, we adopt the eccentric fit for this work since photometry alone cannot constrain eccentricity.
% Instead, we calculate a $3\sigma$ upper limit on the eccentricity to be $e<0.59$. 
% Additionally, we note that the derived eccentricity, $e=0.22^{+0.24}_{-0.12}$, is \textit{not} consistent with zero.
Tables~\ref{tab:fitted_params} and ~\ref{tab:derived_params} show the fitted and derived parameters respectively obtained from the \textcolor{black}{eccentric} fit. We present the fitted transits phase-folded in Fig.~\ref{fig:fitted_transits}.

\subsection{\textcolor{black}{Chromaticity check}} \label{sec:chrom_check}

We run a second (\textcolor{black}{eccentric}) fit of the system with the same priors as in Table~\ref{tab:priors}, however now we allow free dilution ($\mathcal{U}[-1, 1]$) in order to check whether the transits are chromatic (a feature that would be present should the transit signal be produced by an eclipsing binary rather than a planet).
\color{black}
To obtain the depth in each band, we correct the fitted diluted minimum in-transit flux that \texttt{allesfitter} provides for each instrument with the respective derived limb-darkening coefficients and impact parameter. We show these limb-darkening-corrected depths as a function of observation filter in Fig.~\ref{fig:chromaticity}.
The four ground-based depths are mutually consistent, with an inverse-variance weighted mean of $5.685\pm0.166\,$ppt, and $\chi^2=1.05$ for three degrees of freedom ($p=0.789$).
The \textit{TESS} depth, however, lies 5.1$\sigma$ below this mean. Including it yields $\chi^2=26.6$ for four degrees of freedom ($p<10^{-5}$), however we do not consider this indicative of chromaticity: the \textit{TESS} bandpass ($\sim 0.6-1\mu\rm m$) overlaps the \textit{i'} and \textit{z$_{s}$} bands, both of which agree with the \textit{r'} depth, therefore an actual wavelength-dependent depth could not place \textit{TESS} alone below the other bands. Instead, we attribute the offset to under-corrected dilution in the \textit{TESS} photometry.

To test this, we run \texttt{allesfitter} with \textit{TESS} SAP flux, allowing the \textit{TESS} dilution, $D_0$, to vary freely while fixing the dilution to 0 for all ground-based instruments, and compare it to the SPOC value, $D_{0, \rm SPOC}= 1-\mathrm{CROWDSAP} = 0.10$.
Because $R_b/R_\star$ is shared across instruments, the ground-based transits constrain the true depth and break the degeneracy between depth and dilution in the \textit{TESS} data.
We find $D_0=0.29\pm0.04$ (equivalent to a CROWDSAP of 0.71), inconsistent with the SPOC value at $4.8\sigma$.
As the PDCSAP flux has already been corrected for by $D_{0, \rm SPOC}$, this implies a residual dilution in the PDCSAP photometry of $1-(1-D_0)/(1-D_{0, \rm SPOC}) = 0.21\pm0.04$.
In our adopted final fit, we therefore allow free dilution on the \textit{TESS} PDCSAP flux, obtaining $D_{\rm TESS}=0.23_{-0.04}^{+0.05}$, consistent with this expectation and with the depth deficit seen in Fig~\ref{fig:chromaticity}.
Correcting for this dilution increases the PDCSAP depth by a factor of $\sim1.3$, and from $\sim4.5$ to $5.8\,$ppt, in agreement with the ground-based depths.

Our high-resolution imaging (Section~\ref{sec:highres_imaging}) excludes bright companions beyond $\sim0.1\arcsec$, and the ground-based photometry was extracted in uncontaminated apertures; the excess dilution must therefore come from within the much larger \textit{TESS} apertures (21" pixels), but outside the ground-based apertures, thus justifying  zero dilution for the ground-based data.
Considering the other transits and the rest of our validation checks laid out in this paper, we conclude that the transit signal is produced by a planet orbiting TOI-6981.

\color{black}

We also check the stellar density of TOI-6981 produced from the (\textcolor{black}{eccentric} fit) transit analysis \textcolor{black}{($14.56^{+2.29}_{-2.49}\,\rm g\,cm^{-3}$)} with the stellar density as calculated from the stellar mass and radius ($14.82\pm2.16\,\rm g\,cm^{-3}$). We find these to be consistent within $1~\sigma$ indicating that the transiting planet is orbiting the target star.

Finally, we note the potential presence of a star spot visible in some of the photometric observations (e.g., just before mid-transit in the SPECULOOS and \textit{TESS} transits in Figure~\ref{fig:fitted_transits}). While modelling the spot(s) is beyond the scope of this paper, we note that masking out the spotty part of the transit and rerunning the fit provides consistent fitted and derived results.

% fitted parameter table from eccentric, achromtic fit:
\begin{table}
    \centering
    \caption{\textcolor{black}{Fitted parameters obtained from eccentric fit from global analysis of all available photometric data for TOI-6981\,b.}}
    \begin{tabular}{c|c}

    \hline\hline
    Fitted Parameter & Value \\ 
    \hline 
    %\multicolumn{4}{c}{\textit{Fitted parameters}} \\ 
    $R_b / R_\star$ & $0.0755^{+0.0018}_{-0.0015}$ \\ \vspace{0.12cm}
    $(R_\star + R_b) / a_b$ & $0.00760_{-0.00036}^{+0.00048}$\\ 
    $\cos{i_b}$ & $0.0021\pm0.0012$ \\ 
    $T_{0;b}$ (BJD) & $2460349.17655_{-0.00060}^{+0.00074}$ \\ \vspace{0.12cm}
    $P_b$ (d) & $60.778853_{-0.000077}^{+0.000055}$ \\ \vspace{0.12cm}
    $\sqrt{e_b} \cos{\omega_b}$ & $-0.42_{-0.26}^{+0.67}$ \\ 
    $\sqrt{e_b} \sin{\omega_b}$ & $0.13_{-0.17}^{+0.18}$ \\ 
    $D_{0,\rm TESS}$ & $0.23_{-0.04}^{+0.05}$ \\
    $D_{0,\rm all\,others}$ & fixed at 0 \\
    $q_{1; \mathrm{TESS}}$ & $0.2768_{-0.0095}^{+0.0083}$\\
    $q_{2; \mathrm{TESS}}$ & $0.303\pm0.009$\\ 
    $q_{1; \mathrm{r'}}$ & $0.6176_{-0.057}^{+0.0058}$ \\ \vspace{0.12cm}
    $q_{2; \mathrm{r'}}$ & $0.3646_{-0.0084}^{+0.0081}$ \\ \vspace{0.12cm}
    $q_{1; \mathrm{i'}}$ & $0.3702_{-0.0101}^{+0.0079}$ \\ 
    \vspace{0.12cm}
    $q_{2; \mathrm{i'}}$ & $0.2992\pm_{-0.0096}^{+0.0101}$ \\ \vspace{0.12cm}
    $q_{1; \mathrm{z_z}}$ & $0.2779_{-0.0107}^{+0.0086}$ \\ \vspace{0.12cm}
    $q_{2; \mathrm{z_s}}$ & $0.2699_{-0.0087}^{+0.0074}$ \\  \vspace{0.12cm}
    $\ln{\sigma_\mathrm{TESS}}$ & $-5.816\pm0.016$  \\ 
    $\ln{\sigma_\mathrm{ARTEMIS}}$ & $-5.666_{-0.020}^{+0.021}$ \\ 
    $\ln{\sigma_\mathrm{LCO}}$ & $-6.546_{-0.034}^{+0.037}$ \\ 
    $\ln{\sigma_\mathrm{TTT}}$ & $-6.006\pm0.024$\\ 
    $\ln{\sigma_\mathrm{KeplerCam}}$ & $-6.246_{-0.051}^{+0.049}$ \\ 
    \end{tabular}
    \label{tab:fitted_params}
\end{table}

% derived parameter table from eccentric, achromtic fit:
\begin{table}
    \centering
    \caption{\textcolor{black}{Derived parameters obtained from eccentric fit from global analysis of all available photometric data for TOI-6981\,b. Since photometry alone cannot constrain eccentricity, we instead report here the 95 percent confidence upper limit.}}
    %\begin{tabular}{c|c}
    \begin{tabular}{>{\centering\arraybackslash}p{4cm}|c}

    \hline\hline
    Derived Parameter & Value \\ 
    \hline 
    $R_\star/a_\mathrm{b}$ & $0.00706_{-0.00034}^{+0.00045}$ \\ 
    $a_\mathrm{b}/R_\star$ & $141.7_{-8.6}^{+7.1}$ \\ 
    $R_\mathrm{b}/a_\mathrm{b}$ & $0.000535_{-0.000029}^{+0.000033}$ \\ 
    $R_\mathrm{b}$ ($\mathrm{R_{\oplus}}$) & $2.51\pm0.11$ \\ 
    $R_\mathrm{b}$ ($\mathrm{R_{jup}}$) & $0.22\pm0.01$ \\ 
    $a_\mathrm{b}$ ($\mathrm{R_{\odot}}$) & $42.9_{-3.0}^{+2.8}$ \\ 
    $a_\mathrm{b}$ (AU) & $0.1997_{-0.0140}^{+0.0132}$  \\ 
    $i_\mathrm{b}$ (deg) & $89.878\pm0.069$  \\ 
    $e_\mathrm{b}$ & $\leq0.58$ \\
    $b_\mathrm{tra;b}$ & $0.25_{-0.14}^{+0.17}$ \\ 
    $T_\mathrm{tot;b}$ (h) & $3.020_{-0.029}^{+0.038}$ \\ 
    $T_\mathrm{full;b}$ (h) & $2.562_{-0.030}^{+0.021}$ \\ 
    $\rho_\mathrm{\star;b}$ (cgs) & $14.6_{-2.5}^{+2.3}$ \\ 
    $T_\mathrm{eq;b}$ (K) & $174.9_{-5.4}^{+6.3}$ \\ 
    $S_{\rm b}\, (\mathrm{S_\oplus})$ & \textcolor{black}{$0.22\pm0.04$} \\
    
    $u_\mathrm{1; TESS}$ & $0.318\pm0.009$ \\ 
    $u_\mathrm{2; TESS}$ & $0.208\pm0.012$ \\ 
    $u_\mathrm{1; r'}$ & $0.573_{-0.014}^{+0.013}$  \\ 
    $u_\mathrm{2; r'}$ & $0.213\pm0.013$  \\ 
    $u_\mathrm{1; i'}$ & $0.364\pm0.014$\\ 
    $u_\mathrm{2; i'}$ & $0.244\pm_{-0.011}^{+0.012}$ \\ 
    $u_\mathrm{1; z_s}$ & $0.285_{-0.011}^{+0.008}$\\ 
    $u_\mathrm{2; z_s}$ & $0.242_{-0.009}^{+0.010}$  \\ 
    %$\rho_\mathrm{\star; combined}$ (cgs) & $17.1_{-1.9}^{+2.0}$ \\ 

    \end{tabular}
    \label{tab:derived_params}
\end{table}

\section{Discussion and Conclusion} \label{sec:discussion}
\begin{figure*}
    \centering
    \includegraphics[width=0.8\linewidth]{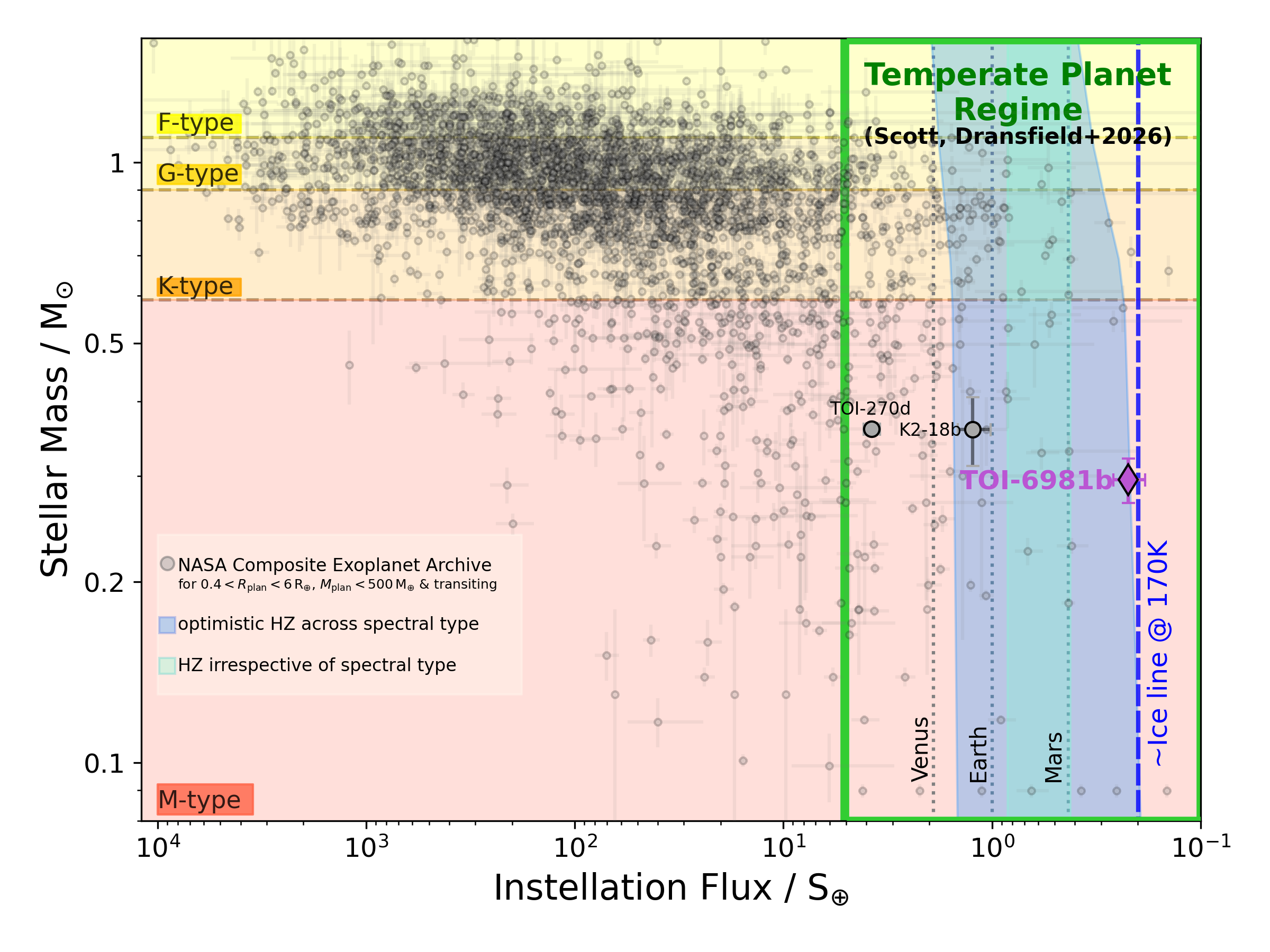}
    \caption{\textcolor{black}{Stellar mass vs instellation flux for all transiting planets (small, grey points) approximately Neptunes or smaller ($\leq6\,\rm R_\oplus$), taken from the NASA Composite Archive. Figure is adapted from \citet{scott2026}. TOI-6981\,b is plotted as a purple diamond, and K2-18\,b \textcolor{black}{and TOI-270\,d as grey circles, all labelled}. The optimistic HZ is depicted by the darker blue shaded region, and the HZ irrespective of stellar type by the cyan shaded region \citep{kopparapu2013}. The water ice line at 170\,K is shown by the blue dashed line, corresponding to an instellation flux of $\sim0.2\,\rm S_\oplus$.
    The temperate planet regime is shown by the green box, ranging from instellation fluxes in the range $0.1\leq S/{\rm S_\oplus} \leq 5$. Background is shaded to indicate different spectral types.}}
    \label{fig:mass_instellation}
\end{figure*}
We present the transiting sub-Neptune TOI-6981\,b, which has a planetary radius of \planetradius, and orbits its host star TOI-6981 with an orbital period of \period. Thanks to the low effective temperature of the star ($T_\mathrm{eff}=3215\pm58\,\rm K$), TOI-6981\,b is rather cold, with an equilibrium temperature of \teq (assuming an Earth-like albedo of 0.3) and an instellation flux of \instellation.
Figure~\ref{fig:mass_instellation} shows TOI-6981\,b in stellar mass - instellation flux space for all transiting planets $\lesssim$ Neptune-size ($<6\,\rm R_\oplus$), adapted from \citet{scott2026}. TOI-6981\,b sits just within the colder edge of the optimistic habitable zone, as defined in \citet{kopparapu2013}, and is within the top five coldest transiting planets $\lesssim$Neptune-sized currently known. 

\subsection{TEMPOS}
TOI-6981\,b is part of the TEMPOS sample \citep[defined in][]{scott2026}, which has the goal of obtaining a catalogue of very precisely ($\leq3\%$) characterised planet radii for temperate planets ($0.1\,\leq  S/\mathrm{S_\oplus} \leq 5$, $T_{\rm eq}\leq400\,\rm K$) orbiting low-mass stars ($T_\mathrm{eff}\leq3400\,\rm K$), and at present is the \textit{coldest} planet in this sample. 
While the precision on the planetary radius (\radiusprecision) does not currently meet the TEMPOS goal, future additional transits and / or better constraints on the stellar parameters will aid in improving this. However it should be noted that due to the long orbital period of TOI-6981\,b, the next \textit{full} transit from the ground is not expected to be until December 2027.

\subsection{Tidal circularisation}
\color{black}

With the currently available data, our global analysis favours an eccentric orbit over a circular one ($\Delta \ln Z =3.1$; Section~\ref{sec:analysis}), with $e\leq0.58$. 
However, photometry constrains eccentricity only indirectly through the transit duration and its consistency with the independently derived stellar density (the photo-eccentric effect; \citealt{dawson2012}), and $e$ remains degenerate with argument of periastron, $\omega$.
To assess whether a non-zero eccentricity is physically plausible, we calculate the tidal circularisation timescale of TOI-6981\,b,

\begin{equation}
    \tau_{\rm circ} = \frac{4}{63} Q' \sqrt{\frac{a^3}{GM_\star}} \bigg(\frac{M_p}{M_\star}\bigg) \bigg(\frac{a}{R_p}\bigg)^{5},
\end{equation}

where $Q'=Q/k_2$ is the modified tidal quality factor of the planet (with $k_2$ being the Love number), $a$ is the semi-major axis, $M_p, M_\star$ are the mass of the planet and the star respectively, and $R_p$ is the radius of the planet \citep{Goldreich1966,rasio1996}.

For sub-Neptunes, we assume a tidal quality factor of $Q\sim10^4$ similar to that of Neptune and Uranus \citep{tittimore1990, zhang2008}, and a Love number value of $k_2\sim0.4$ (typical for gas giants).
We find that TOI-6981\,b would need $7.4\times10^{4}\,$Gyr to circularise, far exceeding the age of the system ($\lesssim1.5\,\rm Gyr$; Section~\ref{sec:TRES}).
Any primordial or dynamically excited eccentricity could therefore persist to the present day, consistent with the eccentric solution preferred by the photometry.
However, in order to properly constrain the eccentricity of TOI-6981\,b's orbit, we would require precise radial velocity data.

% While the available data do prefer a circular orbit at present, we know that photometry alone cannot constrain orbital eccentricity. Thus, we calculate the circularisation timescale for TOI-6981\,b due to the effect of tides using,

% \begin{equation}
%     \tau_{\rm circ} = \frac{4}{63} Q' \sqrt{\frac{a^3}{GM_\star}} \bigg(\frac{M_p}{M_\star}\bigg) \bigg(\frac{a}{R_p}\bigg)^{5},
% \end{equation}

% where $Q'=Q/k_2$ is the modified tidal quality factor of the planet (with $k_2$ being the Love number), $a$ is the semi-major axis, $M_p, M_\star$ are the mass of the planet and the star respectively, and $R_p$ is the radius of the planet \citep{Goldreich1966,rasio1996}. 
% For sub-Neptunes, we assume a tidal quality factor of $Q\sim10^4$ similar to that of Neptune and Uranus \citep{tittimore1990, zhang2008}, and a Love number value of $k_2\sim0.4$ (typical for gas giants). We find that TOI-6981\,b would need $7.4\times10^{4}\,$Gyr to circularise, therefore suggesting the orbit may in fact not be circular. However, in order to properly constrain the eccentricity of TOI-6981\,b's orbit, we would require precise radial velocity data.

\color{black}
\subsection{TOI-6981\,b and the water ice line}

TOI-6981\,b's current location in its orbit places it roughly at the location of the water ice line of its host star, with \teq (see Fig.~\ref{fig:mass_instellation}).
While we are currently unable to directly constrain its bulk density due to the lack of a mass measurement, its sub-Neptune size (\planetradius) is consistent with a volatile-rich interior \citep{rogers2015, fulton2017}. 
Whether this reflects in-situ formation at the water ice line \citep{ormel2010, lambrechts2012} or inwards migration from a much wider orbit \citep{bitsch2015} remains unconstrained without further follow-up.

% As described in Section~\ref{sec:intro}, we define the ice line to be at temperatures of $\sim 150-170\, \rm K$. TOI-6981\,b sits at approximately the ice-line of its host star. 
% While 
% Based on temperature alone, TOI-6981\,b sits just ahead of the general ice line definition with \teq.

% However, we emphasise that without a mass measurement or atmospheric observations, 

% \color{orange}
% \begin{itemize}
%     \item comment on eccentricity and whether that makes it dip in and out?
%     \item what can we say about its formation and migration? if anything? is it worth even speculating without a mass and atmosphere?
% \end{itemize}
% \color{black}

\subsection{Future prospects - planetary mass \& atmospheric studies}

\subsubsection{Prospects for a mass measurement}
In order to properly characterise TOI-6981\,b, a precise mass measurement is required. Since at the time of writing we do not have sufficiently precise RV data of TOI-6981, we instead estimate the mass using the mass-radius relations described in \citet{chenkipping2017}. We calculate the mass to be \planetmass, leading to a density of \planetdensity assuming that TOI-6981\,b is likely a volatile-rich planet with a H/H$_{\rm e}$ envelope as mentioned above. 

Assuming this mass, we calculate the expected RV semi-amplitude to be \textcolor{black}{$K_{\rm pred}=2.5\pm0.3\,\rm m\,s^{-1}$}. Below, we estimate the expected number of RVs required to reach a $5~\sigma$ mass precision assuming this semi-amplitude with two current facilities optimised for precise mass characterisation for planets orbiting M dwarfs, namely MAROON-X \citep{maroonx} and ESPRESSO \citep{espresso}. 

Using the MAROON-X ETC\footnote{MAROON-X exposure time calculator - \url{https://maroon-x-etc.gemini.edu/app}}, we find that for an 1800\,s exposure, assuming typical seeing conditions (airmass\,$=1.5$, cloud cover\,$=50$\%-ile, image quality\,$=70$\%-ile), the expected SNR is $\sim130$, leading to a predicted RV precision of $\sigma_{\rm RV}=1.62\,\rm m\,s^{-1}$. 
% However, we note that jitter is not taken into account in the intrinsic MAROON-X noise estimation, thus this value is likely an under-estimate of the true RV precision we would achieve from these observations.
% Also, observations can benefit from the fact that MAROON-X has two arms (blue and red). While the precision from the data in the blue arm is typically less than in the red due to the wavelength coverage (M dwarfs have better pre
From this, we estimate that \textcolor{black}{17} RV points would be required for a  $5~\sigma$ mass precision \textcolor{black}{(assuming only 2/3\,s of the RVs will be useful)}. 
\color{black}
Similarly for ESPRESSO, we predict an SNR of $\sim23$ and an RV precision of $\sigma_{\rm RV}=2.57\,\rm m\,s^{-1}$ using the ESPRESSO ETC\footnote{ESPRESSO exposure time calculator - \url{https://www.eso.org/observing/etc/bin/gen/form?INS.NAME=ESPRESSO+INS.MODE=spectro}}, requiring 42 RV points for a $5~\sigma$ detection assuming an 1800\,s exposure.

\color{black}
We remind the reader that this is an illustrative example meant to provide an idea about how easy/hard RV follow-up is, \textcolor{black}{in a purely idealistic scenario (i.e., no stellar jitter}).
Other practicalities such as sampling the orbital phase, and stellar activity have been neglected in our exercise, \textcolor{black}{since we do not have the current precision from the TRES RVs to constrain the impact of stellar activity.}

\color{black}

Of course, it is evident from the photometry (see Figure~\ref{fig:tess}) that TOI-6981 is active, with frequent flares occurring within the 27\,d \textit{TESS} sectors. While we do not perform a comprehensive analysis of the stellar activity and flares in this paper, this fact should be considered for future follow-up studies and will likely affect the mass characterisation of TOI-6981\,b, likely implying that more RVs are needed to well-characterise the mass than are calculated above.

While we could assume a stellar activity jitter from TRES from similar stars in the literature, e.g., LP 768-113, (which has a similar mass $M_\star=0.29\,\rm M_{\odot}$, H$\alpha$ equivalent width$=-2.0\,\AA$ \citep{Pass2023b} and an RV jitter of $9.6\,\rm m\,s^{-1}$ \citep{ruh2024}) we see from studies such as \citet{lafarga2021} that RV jitter can be over one order of magnitude different between  stars with a measured H$\alpha$ equivalent width similar to the value of $-2.3\pm0.43\,\AA$ we obtain for TOI-6981 (Section~\ref{sec:TRES}).%, Figure 7 in this paper shows that the RV jitter can span over an order of magnitude.
Including a range of jitters, we find that the calculated number of RV points needed could range from 23 ($1\,\rm m\,s^{-1}$ RV jitter) to 647 ($10\,\rm m\,s^{-1}$ RV jitter) for MAROON-X (and similarly 48-672 for ESPRESSO).%, thus highlighting how without a known stellar activity measurement, an accurate prediction of the number of RVs needed is difficult and uncontainable.

\color{black}

Another consideration is that since the rotation period of TOI-6981 is a similar order of magnitude to the orbital period of TOI-6981\,b ($\sim43\,\rm d$ and $\sim61\,\rm d$), should the radial velocities be affected by stellar activity, it may also be difficult to disentangle rotation vs planetary effects on the RVs, especially if only a small number of observations are obtained.

We do note however that while there could be an impact on the RVs from stellar activity from e.g., spots, broadening of spectral lines should not be an issue thanks to the upper limit on $v\sin i\lesssim3.4\,\rm km\,s^{-1}$ (obtained from the TRES spectra -- see Section~\ref{sec:TRES}). In fact, calculating $v\sin i$ from the photometric rotation period of $43\,\rm d$ gives an tighter upper limit at $v\sin i\lesssim0.36\,\rm km\,s^{-1}$.%, further proving broadening of the spectral lines should not affect the mass characterisation of TOI-6981\,b.

\color{black}

\subsubsection{Prospects for atmospheric characterisation}
\begin{figure}
    \centering
    \includegraphics[width=1\linewidth]{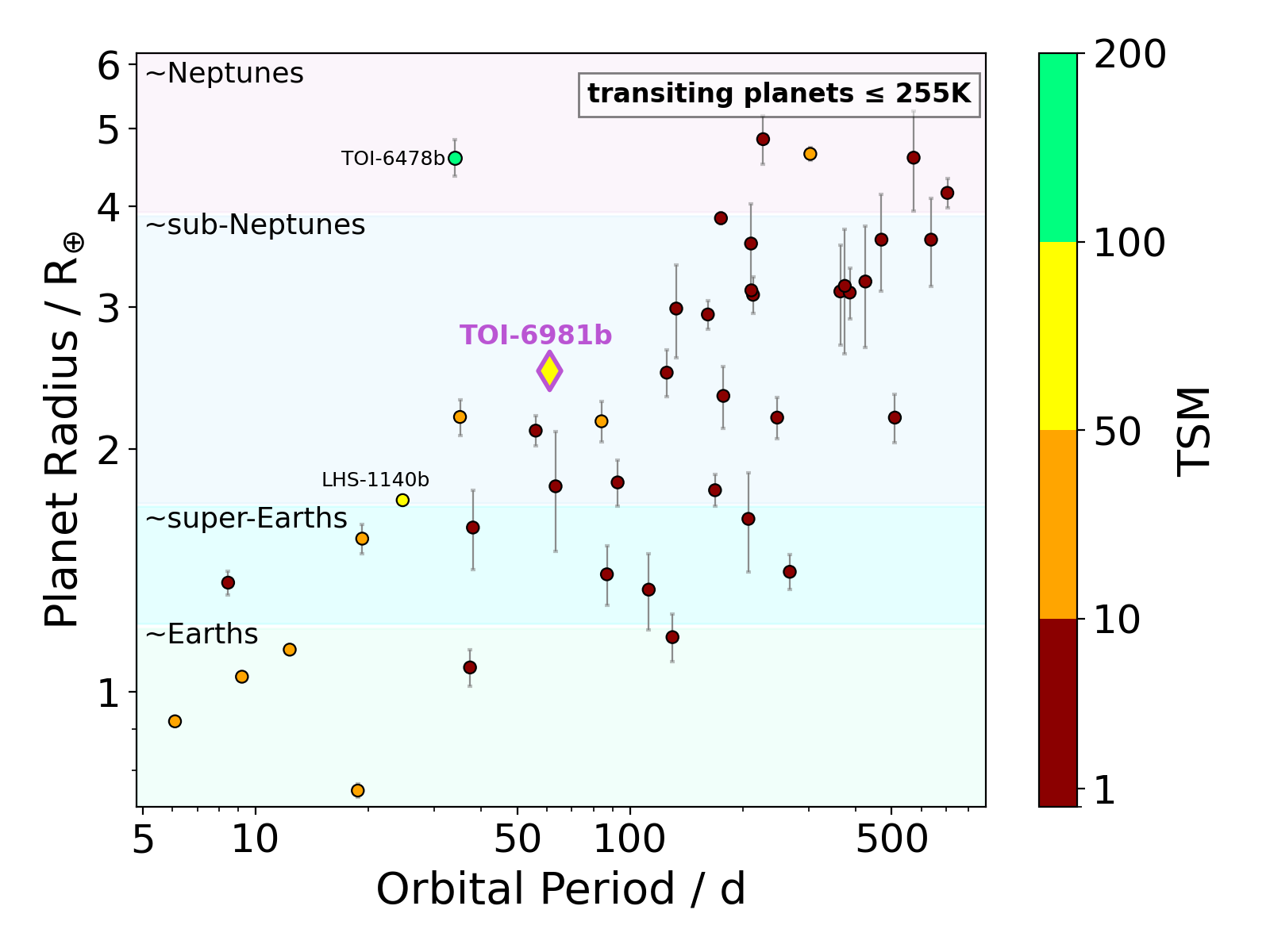}
    \caption{\textcolor{black}{Planet radius vs orbital period for transiting approximately Neptunes or smaller ($\leq6\,\rm R_\oplus$, same as Fig~\ref{fig:mass_instellation}) but only for planets colder than Earth ($\leq$255\,K, Earth's temperature assuming an albedo of 0.3). The colour bar shows the transmission spectroscopy metric \citep{kempton2018}, recalculated for all planets. Planets are approximately split into their planet-types: $\sim$Earths, $\sim$super-Earths, $\sim$sub-Neptunes, $\sim$Neptunes (bottom to top). 
    TOI-6981\,b is depicted by a diamond with a purple outline. Two other key planets, TOI-6478\,b and LHS-1140\,b are labelled for reference. }
    }
    \label{fig:tsm}
\end{figure}
Assuming the mass above (\planetmass), we calculate the transmission spectroscopy metric \citep[TSM;][]{kempton2018}, which is proportional to the expected SNR of transmission spectrum features with \textit{JWST}, to be \tsm. 
When comparing to other cold ($T_{\rm eq}\leq255\,\rm K$), transiting exoplanets (Fig~\ref{fig:tsm}), we find that TOI-6981\,b has the third highest predicted TSM, beaten only by TOI-6478\,b \citep[$\mathrm{TSM}\sim230$;][and part of the HST Observing Program 17414, PI L.Kreidberg\footnote{\url{https://www.stsci.edu/hst/phase2-public/17414.pro}}]{scott2025}, and LHS-1140\,b ($\mathrm{TSM}\sim68$; an already well-studied super-Earth with \textit{JWST} see e.g., Program ID: 6543, PI: C. Cadieux \& R. Doyon, \citep{cadieux2024}; Program ID: 2334, PI: M. Damiano, \citep{damiano2024}; \textcolor{black}{and recently \citealt{Cherubim2026}, who detect helium escaping its atmosphere}).
For consistency, we recalculate all TSMs. For TOI-6478\,b, and assume the upper mass limit of $\sim10\,\rm M_{\oplus}$ as in \citet{scott2025}.
% \citep[$\mathrm{TSM}\sim68$;][an already well-studied super-Earth with \textit{JWST} see e.g.,Program ID: 6543, PI: C. Cadieux \& R. Doyon, \citep{cadieux2024}; Program ID: 2334, PI: M. Damiano, \citep{damiano2024}; \textcolor{red}{and recently \citealt{Cherubim2026}, who detect helium escaping its atmosphere}]{}.

\subsubsection{Prospects for testing the Hycean world hypothesis}
With future atmospheric studies, TOI-6981\,b could provide insight into the currently debated Hycean world hypothesis, which proposes that some sub-Neptune planets can host liquid water oceans beneath massive H$_2$-dominated atmospheres \citep{madhu2021, madhu2023, seager2013}.
With \teq, TOI-6981\,b is too cold to host liquid surface water in the absence of substantial atmospheric greenhouse warming; this makes it a particularly clean test case for the Hycean world hypothesis.
If there is a detection of \textcolor{black}{H$_2$O or a biosignature gas} in the transmission spectrum of the planet \citep{madhu2021}, this would imply a significant greenhouse contribution that is enough to raise the surface temperature above the water melting point, as H$_2$-dominated atmospheres are known to be efficient greenhouses at low temperatures \citep{pierrehumbert2011, seager2013}.
This contrasts warmer proposed Hycean candidates such as K2-18\,b \citep{madhu2023}, whose higher equilibrium temperature ($T_{\mathrm{eq}}\sim280\,\rm K$) means that the presence of liquid water does not necessarily require an atmospheric explanation, and as such can bring forth debate on the interpretation of atmospheric detections.

\color{black}
\subsubsection{Prospects for M dwarf planet habitability}

The properties of TOI-6981\,b also have broader implications for the habitability of planets orbiting M dwarfs.
It is widely debated whether M-dwarf planets can sustain life given the intense XUV (X-ray and extreme ultra-violet) radiation, stellar flares and strong stellar winds that these stars produce which can cause rapid atmospheric escape and erode planet atmospheres \citep{shields2016, france2016, amaral2022}.
This debate has been reinforced by recent \textit{JWST} observations of the inner TRAPPIST-1 planets (b and c), which show evidence for bare rock or very thin atmospheres that are consistent with atmospheric loss \citep{greene2023, zieba2023, ducrot2025, gillon2026}.
However, thanks to the wider orbital period of TOI-6981\,b relative to the classical habitable zone \citep[HZ;][]{kopparapu2013}, its XUV flux exposure is significantly reduced. As well, having a potentially thick H$_2$ atmosphere, as discussed relative to the Hycean world hypothesis above, could further aid in resisting atmospheric stripping from harsh space weather.
Together, these factors suggest that cold, relatively wide-orbiting M dwarf sub-Neptunes could represent a more favourable environment for atmospheric retention compared to closer-in, warmer, classical HZ sub-Neptunes. If liquid water can be maintained, they could potentially broaden the parameter space in which life around M dwarfs may exist.

\color{black}

\subsection{Summary}
In summary, we present TOI-6981\,b, a cold, temperate (\teq) sub-Neptune (\planetradius) orbiting at the ice-line of its M-dwarf host star with an orbital period of just \period. 
We validate the planetary nature of this object using high-resolution imaging, statistical analysis and chromaticity investigations.
Placing TOI-6981\,b in context with the wider planetary population, we see that it is one of the coldest transiting planets $\lesssim$Neptune-sized known, sitting just at the outer edge of the optimistic habitable zone.
For planets colder than Earth, it is the most amenable planet (\textcolor{black}{assuming the predicted mass from mass-radius relations)}  for atmospheric transmission studies that has not already been observed with \textit{HST} or \textit{JWST}, and could provide insight and clarity on the Hycean world debate for sub-Neptunes.

% % conclude:
% In summary...
% \color{black}

\section*{Acknowledgements}
MGS acknowledges support from the UK Science and Technology Facilities Council (STFC) and from a local studentship delivered by the College of Engineering and Physical Sciences of the University of Birmingham.
This paper includes data collected by the TESS mission. Funding for the TESS mission is provided by the NASA's Science Mission Directorate.

This article is based on observations made in the 
Two-meter Twin Telescope (TTT\footnote{\url{http://ttt.iac.es}}) sited at the Teide Observatory of the Instituto 
de Astrofísica de Canarias (IAC), that Light Bridges operates in Tenerife, Canary Islands (Spain). The observation time rights (DTO) used for this research were consumed in the PEI 'TTVARIAT26'.

% LCOGT

This work makes use of observations from the LCOGT network. Part of the LCOGT telescope time was granted by NOIRLab through the Mid-Scale Innovations Program (MSIP). MSIP is funded by NSF.

% ExoFOP

This research has made use of the Exoplanet Follow-up Observation Program (ExoFOP; DOI: 10.26134/ExoFOP5) website, which is operated by the California Institute of Technology, under contract with the National Aeronautics and Space Administration under the Exoplanet Exploration Program.

% TESS

KAC acknowledges support from the TESS mission via subaward s3449 from MIT and NASA grants 80NSSC24K1889 and 80NSSC26K0081.

The ULiege's contribution to SPECULOOS has received funding from the European Research Council under the European Union's Seventh Framework Programme (FP/2007-2013) (grant Agreement n$^\circ$ 336480/SPECULOOS), from the Balzan Prize and Francqui Foundations, from the Belgian Scientific Research Foundation (F.R.S.-FNRS; grant n$^\circ$ T.0109.20), from the University of Liege, and from the ARC grant for Concerted Research Actions financed by the Wallonia-Brussels Federation. 

The Cambridge contribution is supported by a grant from the Simons Foundation (PI Queloz, grant number 327127).

J.d.W. and MIT gratefully acknowledge financial support from the Heising-Simons Foundation, Dr. and Mrs. Colin Masson and Dr. Peter A. Gilman for Artemis, the first telescope of the SPECULOOS network situated in Tenerife, Spain. 

The Bern contribution is supported by the Swiss National Science Foundation (PP00P2-163967, PP00P2-190080 and the National Centre for Competence in Research PlanetS). 

The Birmingham contribution to SPECULOOS has received fund from the European Research Council (ERC) under the European Union's Horizon 2020 research and innovation programme (grant agreement n$^\circ$ 803193/BEBOP), from the MERAC foundation, and from the Science and Technology Facilities Council (STFC; grant n$^\circ$ ST/S00193X/1, ST/W002582/1, and ST/Y001710/1) and from the ERC/UKRI Frontier Research Guarantee programme (EP/Z000327/1/CandY).
 
R.A. acknowledges support from the Spanish Research Agency of the Ministry of Science, Innovation and Universities (AEI-MICIU) under grant PID2023-149439NB-C41.

This article used flash storage and GPU computing resources as Indefeasible Computer Rights (ICRs) being commissioned at the ASTRO POC project that Light Bridges will operate in the Island of Tenerife, Canary Islands (Spain). The ICR were consumed in
the PEI 'TTVARIAT26' with the collaboration of Bechtle and LENOVO.
% Khalid
Funding for KB was provided by the European Union (ERC AdG SUBSTELLAR, GA 101054354).
% BVR
This material is based upon work supported by the National Aeronautics and Space Administration under Agreement No.\ 80NSSC21K0593 for the program ``Alien Earths''.
The results reported herein benefited from collaborations and/or information exchange within NASA’s Nexus for Exoplanet System Science (NExSS) research coordination network sponsored by NASA’s Science Mission Directorate.
This material is based upon work supported by the European Research Council (ERC) Synergy Grant under the European Union’s Horizon 2020 research and innovation program (grant No.\ 101118581---project REVEAL).
% IRTF
Visiting Astronomer at the Infrared Telescope Facility, which is operated by the University of Hawaii under contract 80HQTR24DA010 with the National Aeronautics and Space Administration.
% YGMC, AK and MPM
YGMC, AK, and MPM are partially supported by UNAM PAPIIT-IG101224 and by the Swiss National Science Foundation IZSTZ0\_216537.
Author F.J.P acknowledges financial support from the Severo Ochoa grant CEX2021-001131-S funded by MCIN/AEI/10.13039/501100011033 and Ministerio de Ciencia e Innovación through the project PID2022-137241NB-C43.

MG is F.R.S.-FNRS Research Director.

%%%%%%%%%%%%%%%%%%%%%%%%%%%%%%%%%%%%%%%%%%%%%%%%%%
\section*{Data Availability}
\textit{TESS} data products are available via the MAST portal at \url{https:// mast.stsci.edu/portal/Mashup/Clients/Mast/Portal.html}.
Follow- up observations (photometry, high-resolution imaging data) are available on ExoFOP or on request.



\bibliographystyle{mnras}
\bibliography{references} % if your bibtex file is called example.bib

\vspace{1cm}
\noindent % List of institutions
$^{1}$School of Physics and Astronomy, University of Birmingham, Edgbaston, Birmingham B15 2TT, UK\\
$^{2}$Kavli Institute for Astrophysics and Space Research, Massachusetts Institute of Technology, Cambridge, MA 02139, USA \\
$^{3}$Center for Astrophysics | Harvard \& Smithsonian, 60 Garden Street, Cambridge, MA, 02138, USA \\
$^{4}$Instituto de Astrofísica de Canarias (IAC), Calle Vía Láctea s/n, 38200, La Laguna, Tenerife, Spain\\
$^{5}$Astrobiology Research Unit, Université de Liège, 19C Allée du 6 Août, 4000 Liège, Belgium\\
$^{6}$Department of Earth, Atmospheric and Planetary Science, Massachusetts Institute of Technology, 77 Massachusetts Avenue, Cambridge, MA 02139, USA\\
$^{7}$Department of Astronomy \& Astrophysics, UC San Diego, La Jolla, CA 92039, USA\\
$^{8}$NSF NOIRLab, 950 N. Cherry Ave., Tucson, AZ 85719, USA \\
$^{9}$NASA Ames Research Center, Moffett Field, CA 94035, USA\\
$^{10}$Sternberg Astronomical Institute Lomonosov Moscow State University, Universitetskii prospekt, 13 Moscow, Russia\\
$^{11}$Departamento de Astrofísica, Universidad de La Laguna, Avda. Astrofísico Francisco Sánchez, E-38206 La Laguna, Tenerife, Spain\\
$^{12}$Light Bridges, SL. Observatorio Astron\'omico del Teide, Carretera del Observatorio s/n, E-38500 Guimar, Tenerife, Canarias, Spain\\
$^{13}$Department of Astrophysics, University of Oxford, Denys Wilkinson Build- ing, Keble Road, Oxford OX1 3RH, UK\\
$^{14}$Magdalen College, University of Oxford, Oxford OX1 4AU, UK\\
$^{15}$Kotizarovci Observatory, Sarsoni 90, 51216 Viskovo, Croatia\\
$^{16}$University of Southern Queensland, Centre for Astrophysics, West Street, Toowoomba, QLD 4350, Australia\\
%$^{q}$Physikalisches Institut, University of Bern, Sidlerstrasse 5, 3012, Bern, Switzerland\\
$^{17}$AIM, CEA, CNRS, Université Paris-Saclay, Université de Paris, F-91191 Gif-sur-Yvette, France\\
$^{18}$Universidad Nacional Autónoma de México, Instituto de Astronomía, AP 70-264, Ciudad de México, 04510, México\\
$^{19}$Cavendish Laboratory, JJ Thomson Avenue, Cambridge, CB3 0HE, UK\\
$^{20}$Space Sciences, Technologies and Astrophysics Research (STAR) Institute, Université de Liège, Allée du 6 Août 19C, B-4000 Liège, Belgium \\
$^{21}$Institute for Particle Physics and Astrophysics, ETH Zürich, Wolfgang-Pauli-Strasse 2, 8093 Zürich, Switzerland \\
$^{22}$Instituto de Astrofísica de Andalucía (IAA-CSIC), Glorieta de la Astronomía s/n, 18008 Granada, Spain\\
$^{23}$Max Planck Institute for Astronomy, Königstuhl 17, 69117 Heidelberg, Germany\\
$^{24}$Department of Astronomy \& Space Sciences, Faculty of Science, Ankara University, TR-06100, Ankara, T\"urkiye\\
$^{25}$Ankara University, Astronomy and Space Sciences Research and Application Center (Kreiken Observatory), Incek Blvd., TR-06837, Ahlatlıbel, Ankara, T\"urkiye\\
$^{26}$Center for Space and Habitability, University of Bern, Gesellschaftsstrasse 6, 3012, Bern, Switzerland\\
% $^{1}$School of Physics and Astronomy, University of Birmingham, Edgbaston, Birmingham B15 2TT, UK\\
% $^{2}$\\
% $^{3}$
% $^{xx}$Department of Astronomy \& Astrophysics, UC San Diego, La Jolla, CA 92093, USA\\ 
% $^{5}$Light Bridges, Observatorio Astronómico del Teide. Carretera del Observatorio del Teide, s/n, E-38570 Güímar, Tenerife, Spain \\
% $^{6}$Instituto de Astrofísica de Canarias (IAC), Vía Láctea, s/n, E-38205, La Laguna, Tenerife, Spain \\
% $^{7}$Departamento de Astrofísica, Universidad de La Laguna (ULL), E-38206 La Laguna, Tenerife, Spain \\
% $^{a}$Department of Astronomy \& Space Sciences, Faculty of Science, Ankara University, TR-06100, Ankara, T\"urkiye\\
% $^{b}$Ankara University, Astronomy and Space Sciences Research and Application Center (Kreiken Observatory), Incek Blvd., TR-06837, Ahlatlıbel, Ankara, T\"urkiye\\
% $^{c}$Astrobiology Research Unit, Universit\'e de Li\`ege, All\'ee du 6 Ao\^ut 19C, B-4000 Li\`ege, Belgium

%%%%%%%%%%%%%%%%%%%%%%%%%%%%%%%%%%%%%%%%%%%%%%%%%%

%%%%%%%%%%%%%%%%% APPENDICES %%%%%%%%%%%%%%%%%%%%%

\appendix

\color{black}
\section{\textit{TESS} target pixel file}

\begin{figure}
    \centering
    \includegraphics[width=1\linewidth]{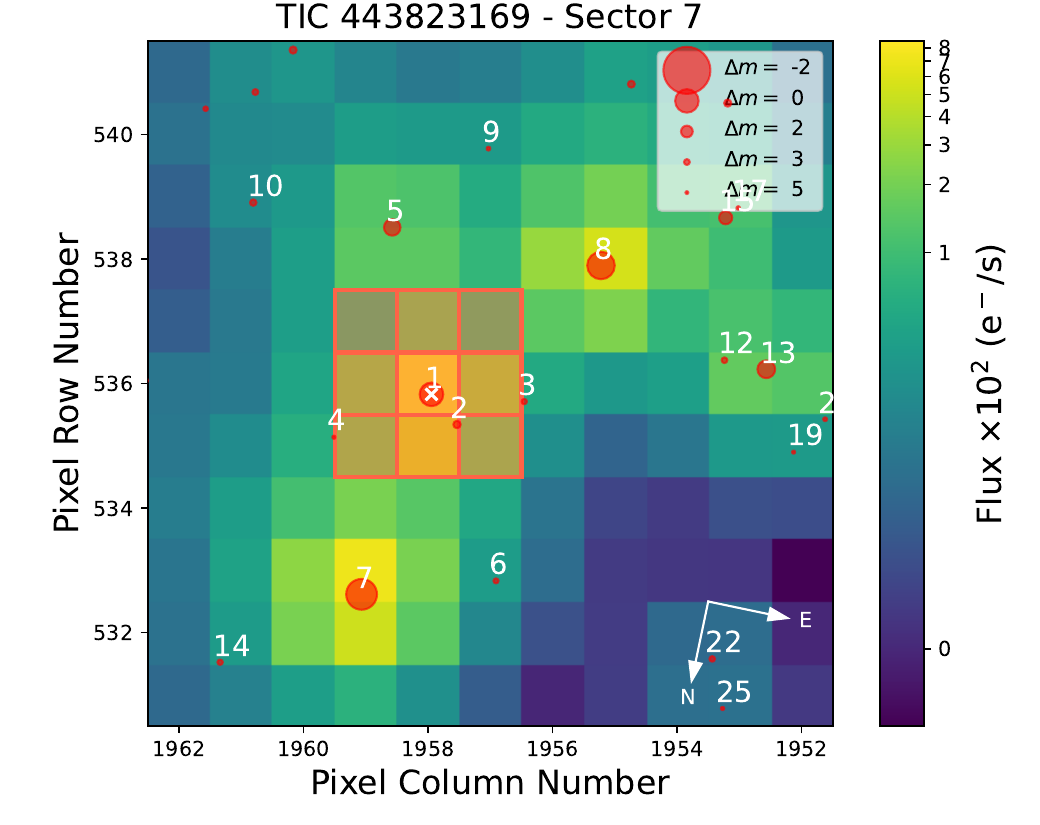}
    \caption{\textcolor{black}{\textit{TESS} target pixel file for TOI-6981 showing the default \textit{TESS} aperture used to extract the photometry. Nearby sources and their magnitudes are highlighted.}}
    \label{fig:tpf}
\end{figure}

\color{black}
\section{Follow-up observations}

\begin{table*} 
\centering
\caption{Summary of ground-based follow-up observations carried out for TOI-6981.}
\begin{tabular}{@{}ccccc@{}}
\midrule \midrule
\multicolumn{5}{c}{\textbf{TOI-6981 Follow-up Observations}}                                              \\ \midrule \midrule
\multicolumn{5}{c}{\textbf{High Resolution Imaging}}                                             \\ %\midrule
\textbf{Observatory} & \textbf{Filter} & \textbf{Date}     & \textbf{Sensitivity Limit} & \textbf{Result}\\ \midrule
Gemini \textcolor{black}{North}      & $562~{\rm nm}$     & 2025 Feb 16 & $\Delta m=5.2$ at $0.5\arcsec$  & No sources detected  \\ 
Gemini \textcolor{black}{North}      & $832~{\rm nm}$     & 2025 Feb 16  & $\Delta m=7.13$ at $0.5\arcsec$  & No sources detected   \\
WIYN & $562~{\rm nm}$ & 2025 Feb 04 & \textcolor{black}{$\Delta m=4.7$ at $0.5\arcsec$} & No sources detected  \\ 
WIYN & $832~{\rm nm}$ & 2025 Feb 04 & \textcolor{black}{$\Delta m=5.7$ at $0.5\arcsec$} & No sources detected  \\ 
SAI & $562~{\rm nm}$ (\textit{I$_{\rm c}$}) & 2024 Nov 14 & $\Delta m=5.7$ at $1\arcsec$ & No sources detected  \\ 
\midrule
\multicolumn{5}{c}{\textbf{Photometric Follow-up}}                                               \\% \midrule
\textbf{Observatory} & \textbf{Filter} & \textbf{Date}     & \textbf{Coverage} & \textbf{Result} \\ \midrule
KeplerCam & {\it Sloan-$i'$} & 2024 Dec 08 & Ingress & Detection \\
SPECULOOS-North/Artemis & {\it Sloan-$r'$} & 2026 Feb 06 & Full & Detection \\
LCO-Teid-1m0 & {\it Sloan-$i'$} & 2026 Feb 06 & Full & Detection \\
TTT-2m0 & {\it $z_s$} & 2026 Feb 06 & Full & Detection \\
\multicolumn{5}{c}{\textbf{Spectroscopic Observations}}                                               \\ %\midrule
\textbf{Instrument} & \textbf{Wavelength Range} & \textbf{Date}     & \textbf{Number of Spectra} & \textbf{Use}\\ \midrule
Shane/Kast     & $450-900~\rm nm$      &  \textcolor{black}{2024 Dec 5} & 1  & Stellar characterisation\\\midrule
IRTF/SpeX     & $800-2420~\rm nm$      &  \textcolor{black}{2024 Dec 12} & 1  & Stellar characterisation\\\midrule

 &  & 2025 Oct 26 &  &  \\
TRES & $390-910~\rm nm$ & 2025 Nov 25 & 3 & False positive identification \\
 &  & 2026 Feb 22 &  &  \\\midrule
\end{tabular}
\label{tab:followup_6981}
\end{table*}

\begin{figure}
    \centering
    \includegraphics[width=1\linewidth]{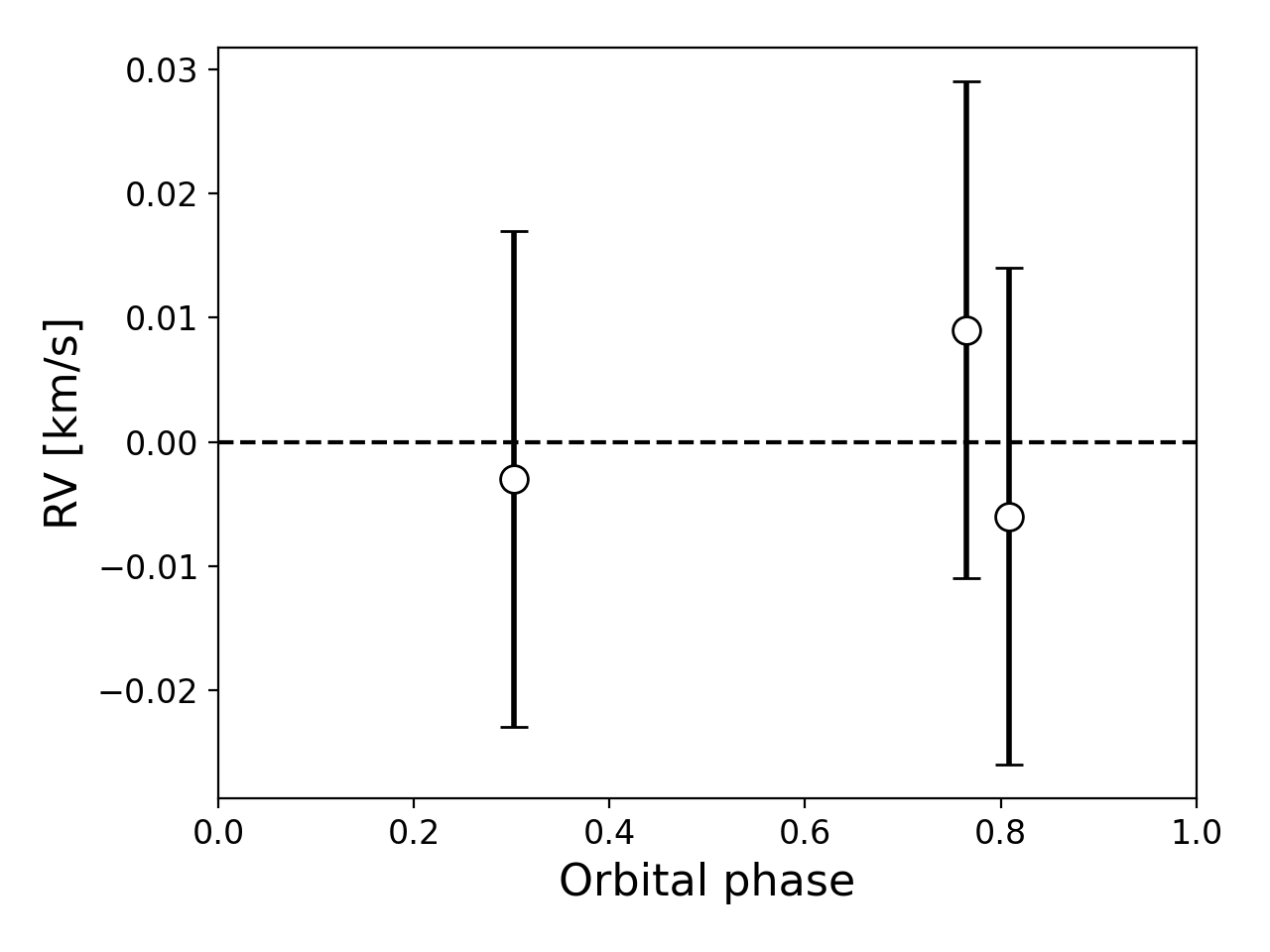}
    \caption{TRES RVs phase-folded on TOI-6981\,b's orbital period ($60.77\,\rm d$).}
    \label{fig:tres}
\end{figure}

\section{\textit{TESS} photometry}

\begin{figure*}
    \centering
    \includegraphics[width=1\linewidth]{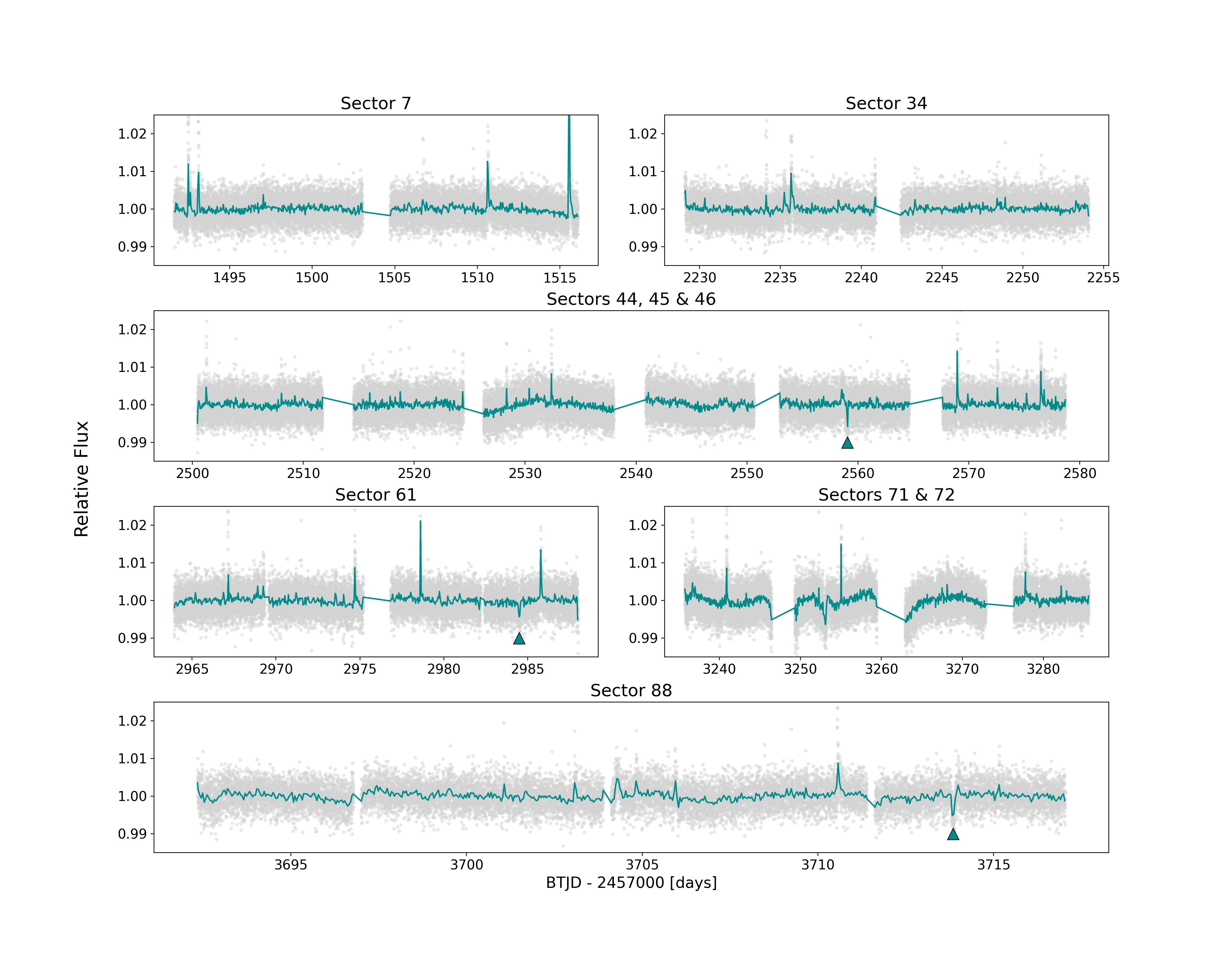}
    \vspace{-1.2cm}
    \caption{120\,s cadence \textit{TESS} data for TOI-6981, with sectors labelled at the top of each panel. The grey points show the two minute exposures and the solid teal line shows this flux binned to one hour. The teal triangles indicate the transits of TOI-6981.01.}
    \label{fig:tess}
\end{figure*}

\section{Archival Imaging}

\begin{figure*}
    \centering
    \includegraphics[width=1\linewidth]{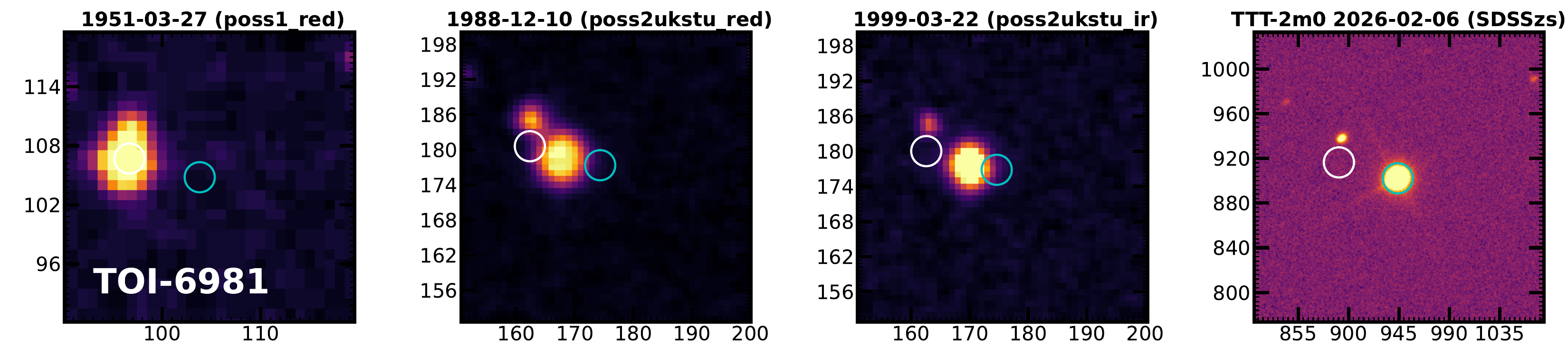}
    \vspace{-0.5cm}
    \caption{Archival data for TOI-6981. White circles show the position of the star in the earliest image, and blue circles show the position of the star in the most recent images. From 1951 to 2026, the target was shifted by 12.7$\arcsec$. No background objects are blending within the current position of TOI-6981.}
    \label{fig:archival}
\end{figure*}

\section{High-Resolution Spectroscopy}

\subsection{Zorro}

\begin{figure}
    \centering
    \includegraphics[width=1\linewidth]{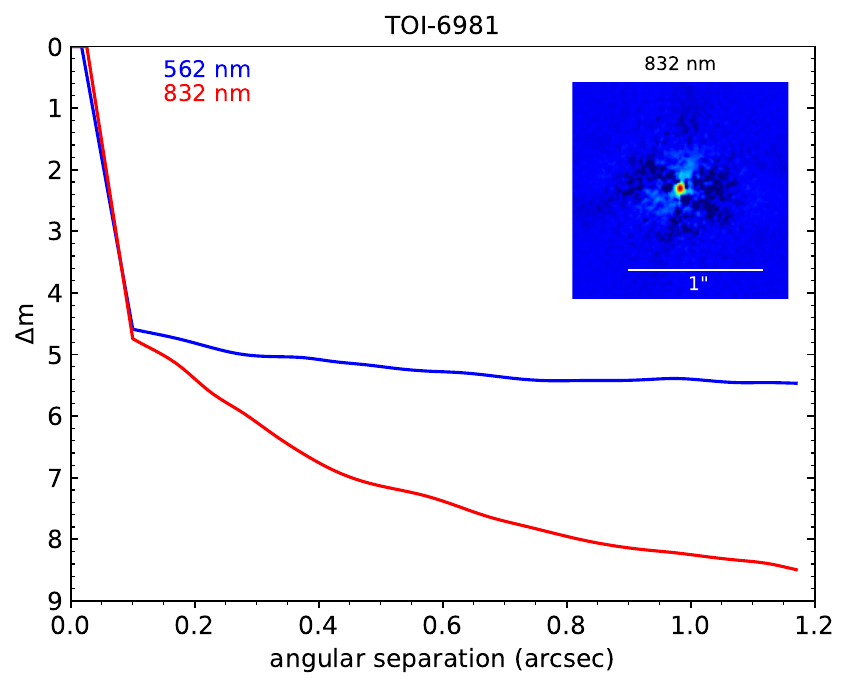}
    \caption{$5\sigma$ magnitude contrast curves in both filters as a function of the angular separation out to 1.2 arcsec, obtained from Zorro. The inset shows the reconstructed 832 nm image of TOI-6981 with a 1 arcsec scale bar. TOI-6981 was found to have no close companions from the diffraction limit (0.02”) out to 1.2 arcsec to within the contrast levels achieved.}
    \label{fig:zorro}
\end{figure}

\subsection{WIYN/NESSI}

\begin{figure}
    \centering
    \includegraphics[width=1\linewidth]{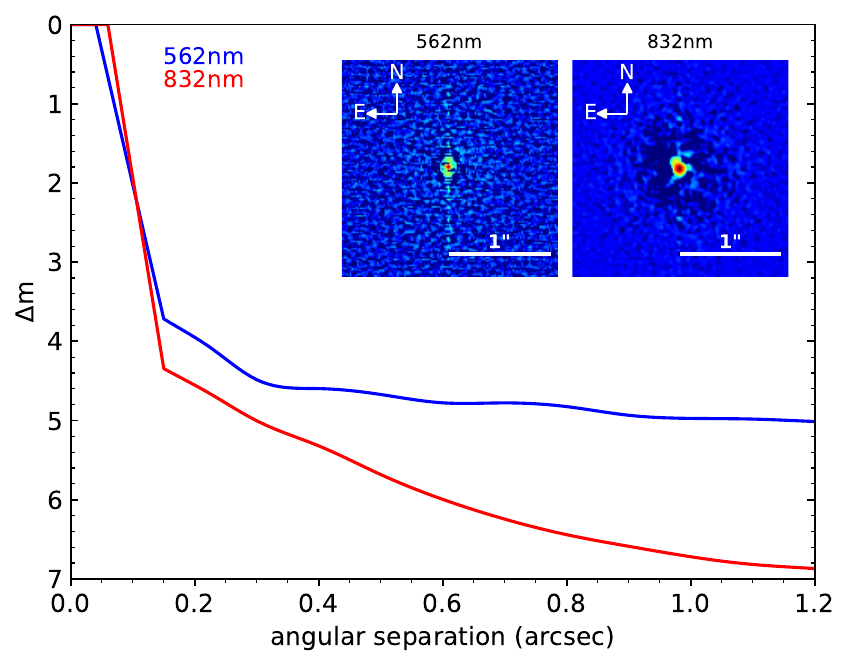}
    \caption{Contrast curves centered on TOI-6981 at 562~nm (blue) and 832~nm (red) obtained from NESSI speckle imaging. Reconstructed images are shown as insets to the plot with wavelengths labelled.
}
    \label{fig:nessi}
\end{figure}

% figure caption: Contrast curves centered on TOI-6981 at 562~nm (blue) and 832~nm (red) obtained from NESSI speckle imaging. Reconstructed images are shown as insets to the plot with wavelengths labelled.

\subsection{SAI}
\begin{figure}
    \centering
    \includegraphics[width=1\linewidth]{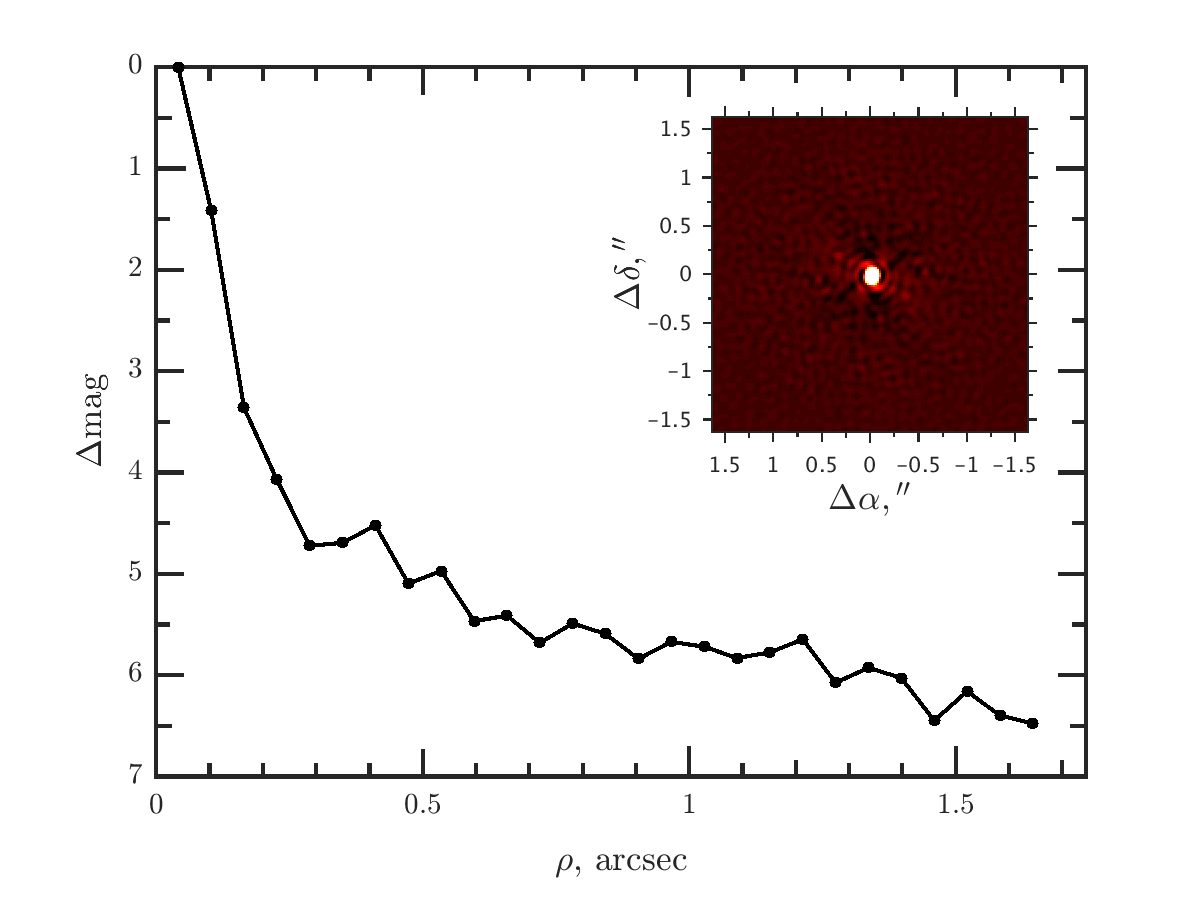}
    \caption{Secondary component detection limits obtained using speckle imaging at speckle polarimeter on the SAI-2.5m telescope. Inset shows autocorrelation function of the object.
    }
    \label{fig:SAI_speckle}
\end{figure}
% zorro, wiyn, sai figures.

\section{Priors for global photometric analysis}
\begin{table} 
\centering
\caption{\textcolor{black}{Priors used in eccentric fit for TOI-6981\,b}}
\begin{tabular}{@{}ccccc@{}} 
\midrule \midrule
Parameter & Fit Parameterization \\ 
\hline
$R_b / R_\star$ & $\mathcal{U}(0.04, 0.14)$ \\
$(R_\star + R_b) / a_b$ & $\mathcal{U}(0.004, 0.012)$ \\
$\cos{i_b}$ &  $\mathcal{U}(0.0, 0.05)$\\
$T_{0;b}$ ($\mathrm{BJD}$) &  $\mathcal{U}(2459559.01, 2459559.08)$\\
$P_b$ ($\mathrm{d}$) & $\mathcal{U}(60.73, 60.8)$ \\
$\sqrt{e_b} \cos{\omega_b}$ & $\mathcal{U}(-1, 1)$ \\
$\sqrt{e_b} \sin{\omega_b}$ & $\mathcal{U}(-1, 1)$ \\
$D_{0,\rm TESS}$ & $\mathcal{U}(-1, 1)$ \\
$D_{0,\rm all\,others}$ & fixed to 0 \\ 
$q_{1; \mathrm{TESS}}$ & $\mathcal{N}_{0,1}(0.278, 0.01)$ \\
$q_{2; \mathrm{TESS}}$ & $\mathcal{N}_{0,1}(0.303, 0.01)$ \\
$q_{1; \mathrm{{r'}}}$ & $\mathcal{N}_{0,1}(0.620, 0.01)$ \\
$q_{2; \mathrm{{r'}}}$ & $\mathcal{N}_{0,1}(0.368, 0.01)$ \\
$q_{1; \mathrm{{i'}}}$ &  $\mathcal{N}_{0,1}(0.369, 0.01)$ \\
$q_{2; \mathrm{{i'}}}$ &  $\mathcal{N}_{0,1}(0.300, 0.01)$\\
$q_{1; \mathrm{{z_s}}}$ & $\mathcal{N}_{0,1}(0.279, 0.01)$  \\
$q_{2; \mathrm{{z_s}}}$ & $\mathcal{N}_{0,1}(0.273, 0.01)$ \\
$\ln{\sigma_\mathrm{all}}$ & $\mathcal{U}(-30, 1)$ \\
\end{tabular} \label{tab:priors}
\end{table}
\color{black}

%%%%%%%%%%%%%%%%%%%%%%%%%%%%%%%%%%%%%%%%%%%%%%%%%%

% Don't change these lines
\bsp	% typesetting comment
\label{lastpage}
\end{document}